\documentclass[11pt,a4paper]{article}
\usepackage[english]{babel}
\usepackage[utf8]{inputenc}
\usepackage{epsf}
\usepackage{cite}
\usepackage{graphicx}
\usepackage{subcaption} % allows subfigures
\usepackage{cancel}
\usepackage{verbatim} % allows block comments

\usepackage{amsmath}
\usepackage{amssymb}
\usepackage{mathrsfs}
\usepackage{multirow}
\usepackage{mathbbol}
\usepackage{slashed}
\usepackage{amsfonts}
\usepackage{bbding}
\usepackage{array}
\usepackage[
colorlinks=true
,urlcolor=black
,anchorcolor=black
,citecolor=black
,filecolor=black
,linkcolor=black
,menucolor=black
,linktocpage=true
,pdfproducer=medialab
,pdfa=true
]{hyperref}
\usepackage{color} % colored text

\usepackage[left=2cm,right=2cm,top=2cm,bottom=2cm]{geometry}
\usepackage{tikz}
\usetikzlibrary{shapes,arrows,positioning,automata,backgrounds,calc,er,patterns}
\usepackage[compat=1.1.0]{tikz-feynman}

\vspace*{1cm}
\allowdisplaybreaks
\vspace*{1cm}
\allowdisplaybreaks
\definecolor{rose}{rgb}{0.5,0.0,0.1}
\definecolor{green2}{rgb}{0,0.6,0.1}
\definecolor{green3}{rgb}{0,0.4,0.2}

\begin{document}

\begin{center}
\vspace*{1mm}
%%%%%%%%%%%%%%%%%%%%%
\vspace{1.3cm}

\bigskip 
\mathversion{bold}
{\Large\bf 
 Thermal effects in freeze-in neutrino dark mater production}

\mathversion{normal}

\vspace*{1.2cm}

{\bf A.~Abada $^{a}$, G.~Arcadi $^{b,c}$,  M.~Lucente $^{d,e}$, G.~Piazza $^{a}$,
 and S.~Rosauro-Alcaraz $^{a}$}

\vspace*{.5cm}
$^{a}$ P\^ole Th\'eorie, Laboratoire de Physique des 2 Infinis Irène Joliot Curie (UMR 9012), \\
CNRS/IN2P3,
15 Rue Georges Clemenceau, 91400 Orsay, France

\vspace*{.2cm}
$^{b}$  Dipartimento di Scienze Matematiche e Informatiche, Scienze Fisiche e Scienze della Terra, \\ Universita degli Studi di Messina, Via Ferdinando Stagno d'Alcontres 31, I-98166 Messina, Italy

\vspace*{.2cm}
$^{c}$ INFN Sezione di Catania, Via Santa Sofia 64, I-95123 Catania, Italy

\vspace*{.2cm}
$^{d}$ Dipartimento di Fisica e Astronomia, Universit\`a di Bologna, via Irnerio 46, 40126 Bologna, Italy

\vspace*{.2cm}

$^{e}$ INFN, Sezione di Bologna, viale Berti Pichat 6/2, 40127, Bologna, Italy

\end{center}

\vspace*{5mm}
\begin{abstract}

\noindent

We present a detailed study of the production of dark matter in the form of a sterile neutrino via freeze-in from decays of heavy right-handed neutrinos.
Our treatment accounts for thermal effects in the effective couplings, generated via neutrino mixing, of the new heavy neutrinos with the Standard Model gauge and Higgs bosons 
and can be applied to several low-energy fermion seesaw scenarios featuring heavy neutrinos in thermal equilibrium with the primordial plasma.
We find that the production of dark matter is not as suppressed as to what is found when considering only Standard Model gauge interactions. 
Our study shows that  the freeze-in dark matter production could  be efficient.

\end{abstract}
\vspace*{10mm}

\newpage

\section{Introduction}

The most significant experimental and observational evidences for physics beyond the Standard Model (SM) are the lack of a dark matter (DM) candidate and the unknown origin of neutrino masses (and of the lepton mixing).
The existence of sterile neutrinos allows to accommodate both these puzzles in a rather straightforward way. Heavy neutral leptons (like right-handed neutrinos), singlets under the Standard Model gauge group, are the most popular option for the generation of the light neutrino masses via different variants of the seesaw mechanism. In this work, we will focus on low-scale seesaw scenarios, i.e. we will assume that the heavy neutral leptons (HNL) responsible for neutrino mass genertion have masses at a scale not exceeding few TeVs. Within this framework it is, moreover, possible to accommodate the masses and couplings of some of these new states to make them stable on cosmological scales and, consequently, be potential DM candidates. Such requirement impacts, in particular, the mass scale of the sterile neutrino DM, favoured to be $\mathcal{O}(1-100)$~keV.

These HNLs mix with the active neutrinos via Yukawa couplings to the SM Higgs boson. In the case of the DM, such mixing is responsible of the production process dubbed Dodelson-Widrow mechanism (DW)~\cite{Dodelson:1993je}. Although this mechanism works for a wide range of masses, it is severely constrained by a variety of observational evidences. First of all, X-ray searches for the decay of the sterile neutrino to an active one and a photon constrain, as function of the mass, the coupling of the sterile neutrinos to the active ones and provide an upper bound, of the order of tens of keV, to the mass of the DM candidate~\cite{Boyarsky:2018tvu}. Very light masses, up to few keV, are instead ruled out by the so called Tremaine-Gunn bound and by structure formation consideration as, for example, phase-space density~\cite{Tremaine:1979we,Boyarsky:2008ju} and the number of satellites of the Milky-Way~\cite{Boyarsky:2012rt,Horiuchi:2013noa,Dekker:2021scf}. For masses above 2~keV, another constraint holds from Lyman-$\alpha$ forest data analysis
(from which it is possible to infer the spectrum of matter density fluctuations, determined by the DM properties). However, this constraint is strongly model dependent and the bounds are related to the warm dark matter (WDM) production mechanism, and to which extent this mechanism contributes to the total DM abundance. It was shown in~\cite{Abada:2014zra} that, once all these constraints are accounted for, the DW mechanism can only account for at most $\sim 50\%$ of the total DM abundance. A possible way out would be guaranteed by the presence of a lepton asymmetry (in such a case, the DM production mechanism is dubbed Shi-Fuller~\cite{Shi:1998km}). This would lead anyway to a very constrained parameter space~\cite{Laine:2008pg,Shaposhnikov:2008pf} unless further ingredients are added to the theory. 

Another economical production mechanism for DM would be provided by the decay, \`{a}-la-freeze-in~\cite{McDonald:2001vt,Chu:2011be,Chu:2013jja,Blennow:2013jba,Klasen:2013ypa}, of heavy right-handed neutrinos. The effectiveness of such mechanism was studied in~\cite{Abada:2014zra}, in the framework of a neutrino model based on the inverse seesaw (ISS)~\cite{Abada:2014vea}, as well as within the Type-I seesaw mechanism~\cite{Lucente:2021har}. In the ISS framework the spectrum of the SM in enlarged by the inclusion of right-handed (i.e. featuring in the interaction basis Yukawa coupling with the SM Higgs) and sterile (i.e. no Yukawa couplings with the Higgs present in the interaction basis) heavy neutrino states. The mass spectrum of the theory features the three light massive neutrinos, and a series of pairs of heavy pseudo-Dirac neutrinos. In the case in which the number of sterile neutrino fields is greater than the number of right-handed ones, additional physical states, with masses lying at intermediate values between the light neutrinos and heavy pseudo-Dirac pairs, are present~\cite{Abada:2014vea}. The most minimal model of this type, complying with neutrino oscillation data, is the so called ISS(2,3) (2 right-handed neutrinos and 3 sterile neutrinos). 
In this case, besides the two pairs of pseudo-Dirac heavy states, an intermediate state (with mass  $m_{DM}\sim \mathcal{O}(\text{keV})$) arises, above the light neutrino masses, providing a good DM candidate~\cite{Abada:2014vea}.\\

In this work we will revisit the studies in Ref.~\cite{Abada:2014zra, Lucente:2021har} on the possibility that a sterile neutrino produced from the decay of heavier HNL accounts, at least partially, for the dark matter component of the Universe. In the ISS mechanism, the Yukawa couplings  are generally  large enough such that the HNL states thermalise in the early Universe, while the light sterile state does not; this enables the freeze-in production of DM via the decay of a heavy HNL state into it and the Higgs boson. The freeze-in mechanism ensures an efficient DM production for a sufficiently suppressed mixing with the active states to evade observational constraints~\cite{Abada:2017ieq,Lucente:2021har}. However, as recently pointed out in~\cite{Fernandez-Martinez:2021ypo}, following a seminal work presented in Ref.~\cite{Lello:2016rvl}, the mixing angle between active and sterile neutrinos is subject, in the early Universe, to potentially large thermal corrections. 
%Indeed, at high temperature, when the latter processes take place, being interaction states, the neutrinos are produced as flavour states in a coherent superposition of the mass states. Being the decoherence length of $\nu_{\rm DM}$ potentially larger than the mean free path of the active neutrinos at high temperature, the corresponding mass state is projected back to the flavour basis before it can decohere. 
Thermal masses are indeed relevant at the time of DM production through freeze-in, and dominate over the keV mass of the neutrino DM considered here, such that the effective light neutrinos propagating in the medium are found to be approximately the flavour ones. %, since thermal masses are generated from interactions and thus are diagonal in flavour basis, the effective active-sterile mixing angle is drastically suppressed at high temperatures 
This renders, for example, freeze-in production via $W$ and $Z$ decays not capable of accounting for the correct amount of DM~\cite{Lello:2016rvl}\footnote{This suppression was not taken into account in~\cite{Datta:2021elq}, implying a final DM abundance many orders of magnitude smaller than the estimate given there, as demonstrated in Ref.~\cite{Lello:2016rvl} and which will be reproduced in the following.}.

We extend the study conducted in~\cite{Lello:2016rvl} to realistic seesaw scenarios, including, in addition to the DM state, additional heavier RH neutrinos (with masses of the order of a few hundreds GeV) with large enough Yukawa interactions to be in thermal equilibrium. These models can generally account for the massive nature of active neutrinos and lepton flavour mixing. These RH neutrinos can then decay into the keV DM and a Higgs boson, typically at temperatures of the order of their masses. Our present work also differs form the one in Ref.~\cite{Lello:2016rvl} by the fact that all the sterile neutrinos, including the DM state, are of Majorana nature. This scenario will ease the embedding of a low-scale fermionic seesaw rendering the framework more realistic and ultimately UV complete. In our analysis, we will consider both the type-I and inverse seesaw mechanisms as benchmark scenarios to illustrate our findings. We focus on production through active-heavy neutrino mixing, thus studying temperatures below the electroweak (EW) phase transition, $T_{SSB}\sim 160$~GeV~\cite{DOnofrio:2014rug}, which we take in the following to be instantaneous. Moreover, in order to be able to make a comparison with previous estimations of the keV neutrino production rate, we only take into account the production through two-body decays. Note however that potentially important contributions to the production rate can arise from $2\rightarrow2$ scatterings as well as from the Landau-Pomeranchuk-Migdal (LPM) effect at high temperatures~\cite{Besak:2012qm,Ghiglieri:2016xye,Ghiglieri:2021vcq}, but we leave the inclusion of such effects, and thus the study of the production rate above the EW crossover, for future work.

This work is organised as follows: the framework and the dark matter production are discussed in Section~\ref{Sec:framework}. Section~\ref{Sec:self-energy} details the  formalism we use and the strategy to evaluate thermal corrections taking into account the Higgs and gauge boson contributions to the neutrino self energy for a general case. We show in Section~\ref{Sec:comparison} how we compare in the case with one light neutrino and one sterile state with the results of earlier work in~\cite{Lello:2016rvl}, finding qualitatively similar conclusions. The analysis for dark matter production below the EW crossover is thoroughly conducted in Section~\ref{Sec:analysis}, first without HNL, and then in the realistic case with two heavy neutrinos in addition to the dark matter candidate. 
Our findings are summarised in Section~\ref{sec:conclusions}. Further technical details of the computations are collected in Appendix~\ref{app:regions_support} and \ref{app:integrals}, while we devote Appendix~\ref{app:scalar-production} to show DM production in an example with a larger dark sector including a new scalar, not suffering from the aformentioned mixing suppression, for completeness.%, we consider the case where the dark sector is comprised of other species and propose an example of production rate with new scalars to complete the dark matter abundance in the form of a sterile neutrino via freeze-in from decays of heavy right-handed neutrinos.

\section{Common origin
for neutrino masses and DM}\label{Sec:framework}

\subsection{Neutrino masses from approximate lepton number symmetry}
Among the several ways to explain the lightness of neutrino masses, the type-I seesaw mechanism~\cite{Minkowski:1977sc, Yanagida:1979as, Glashow:1979nm, Mohapatra:1979ia, GellMann:1980vs, Schechter:1980gr, Schechter:1981cv} provides a simple one by introducing a number of RH neutrinos\footnote{In order to explain oscillation data with the type-I seesaw, it is necessary to introduce at least two RH neutrinos~\cite{Donini:2011jh}.}. On general grounds, given that these RH neutrinos are complete singlets under the SM gauge group, their introduction allows not only for the usual Yukawa coupling to the SM Higgs boson, as for the other fermions of the SM, but also for a Majorana mass term, $M_N$, which breaks lepton number by two units. 
The SM Lagrangian would thus be enlarged with the following lagrangian
\begin{equation}
    \mathcal{L}\supset -\bar{L}_L \tilde{\Phi} Y_{\nu}N_R-\frac{1}{2}\bar{N}_R^c M_N N_R+h.c.,
\end{equation}
where $\Phi$ ($\tilde{\Phi}=i\sigma_2 \Phi^*$) and $L_L$ are the Higgs and lepton doublets, respectively, and $N_R$ the RH neutrinos. Both the Majorana mass $M_N$ and the Yukawa couplings $Y_{\nu}$ are matrices in flavour space. After spontaneous symmetry breaking (SSB), the Higgs develops a vacuum expectation value, $v_H$, and generates a Dirac mass term between SM (active) neutrinos and the RH ones, $m_D\equiv v_HY_{\nu}/\sqrt{2}$. We can work, without loss of generality, in a basis in which the matrix $M_N$ is real and diagonal.
In the seesaw limit, where the scale of the Majorana mass terms is much larger than the Dirac  ones (so that it is possible to define the expansion parameter 
$\Theta = m_D M_N^{-1}$),  
 the light neutrino mass matrix results at leading order from the expression 
\begin{equation}\label{eq:mnu_SS}
 m_{\nu}\simeq -m_D^T M_N^{-1} m_D. 
\end{equation}
In this case, the active-heavy neutrino mixing, which would change leptonic weak interactions, is given at leading order by $\theta \simeq m_D^* M_N^{-1}$ and is in general suppressed due to the relative high-scale of $M_N$. The phenomenological impact of the RH neutrinos can be important if they are not excessively heavy, and have sizeable mixings $\theta$.
In the so-called vanilla seesaw, where no special relation is present in the sub-matrices $M_N$ and $Y_\nu$, the active-sterile mixing is strongly suppressed by the ratio of mass scales between active and sterile neutrinos, {$|\theta| \lesssim \sqrt{m_\nu / M_N } \lesssim 10^{-5} \sqrt{\text{GeV}/M_N}$} resulting, for Majorana masses above the GeV scale, to mixings that are way beyond the reach of current experiments. On the other hand it is important to notice that, being Eq.(\ref{eq:mnu_SS}) a matrix relation, it is perfectly reasonable to make it reproduce current data on neutrino masses and mixing while having an active-sterile mixing that does not follow the vanilla scaling. This may be due to accidental cancellations in the matrix equation or, more interestingly, to an underlying symmetry of the theory. In particular, there exists a one-to-one correspondence between a global lepton number symmetry and massless active neutrinos in the SM extended with fermionic singlets~\cite{Moffat:2017feq} implying that the mass scale of active neutrinos is suppressed (and protected from large radiative corrections) if an approximate lepton number symmetry is present in the theory, even in the presence of sizeable Yukawa couplings. Such fermionic singlet extensions of the SM can feature new physics signals strong enough to be testable in current experiments, while correctly reproducing neutrino oscillation data~\cite{Abada:2007ux}.

Within the type-I seesaw, a convenient parametrisation to effectively explore the parameter space of the model is the Casas-Ibarra (CI) one~\cite{Casas:2001sr}, in which the Dirac mass matrix depends on the low-energy neutrino oscillation parameters as
\begin{equation}\label{eq:CI}
	m_D = - i\ U_\mathrm{PMNS}^* \sqrt{\hat{m}}\ R\ \sqrt{M},
\end{equation}
where $U_\mathrm{PMNS}$ is the lepton mixing matrix, $\hat{m}$ a diagonal matrix containing the masses of the light active neutrinos, $M$ an analogous matrix for the heavy neutrino masses and $R$ is an orthogonal matrix which can be parametrised by  three complex angles\footnote{This is specific to the case with 3 RH neutrinos, one of which we will identify with DM in the following.} $\omega_{ij}$, $R = V_{23} V_{13} V_{12}$, where 
\begin{equation}\hspace*{-0.2cm}
V_{12} = \hspace*{-0.1cm}\left(
	\hspace*{-0.15cm}\begin{array}{ccc}
	\cos\omega_{12} & \sin\omega_{12}& 0 \\
	-\sin\omega_{12} & \cos\omega_{12} & 0 \\
	0 & 0 & 1
	\end{array}
	\hspace*{-0.1cm}\right),
 V_{13} = \hspace*{-0.1cm}\left(\hspace*{-0.15cm}
	\begin{array}{ccc}
	\cos\omega_{13} & 0 & \sin\omega_{13} \\
    0 & 1 & 0 \\
	-\sin\omega_{13} & 0 & \cos\omega_{13} \\
	\end{array}
	\hspace*{-0.1cm}\right),
 V_{23} = \hspace*{-0.1cm}\left(\hspace*{-0.1cm}
	\begin{array}{ccc}
    1 & 0 & 0\\
	0 & \cos\omega_{23} & \sin\omega_{23} \\
	0 & -\sin\omega_{23} & \cos\omega_{23} \\
	\end{array}
	\hspace*{-0.15cm}\right).\label{eq:complexangles}
\end{equation}
The CI parametrisation in Eq.~(\ref{eq:CI}) guarantees to recover the input values of $U_\mathrm{PMNS}$ and $\hat{m}$ from the diagonalisation of the full Lagrangian in Eq.~(\ref{eq:mnu_SS}), as long as the seesaw hypothesis, $\Theta = m_D M_N^{-1} \ll 1$, is satisfied. This means that the imaginary parts of the angles $\omega_{ij}$ in Eq.~(\ref{eq:complexangles}) cannot be too large, as the Yukawa couplings grow exponentially with them, although it is evident that the specific upper bound is also a function of the matrix $M$.
In the framework of the CI parametrisation, a lepton-number conserving scenario is recovered in the limit $\omega_{12}, \omega_{23} \rightarrow 0$, $\omega_{23} \rightarrow i \infty$, $M_1 \rightarrow 0$ and $M_2 \rightarrow M_3$: in this case the mass spectrum of the theory features two massless particles and two massive Dirac states, thus mediating only lepton number-conserving interactions. Such exact limit clearly does not reproduce the observed neutrino data, while a perturbed scenario with an approximate lepton number symmetry (like in for instance the case where $|\omega_{12}|, |\omega_{23}| \ll 1$, $\mathrm{Im}\omega_{23} \gg 1$ and  $M_1 \ll M_{2,3}$) is phenomenologically viable, while featuring a relatively low new-physics scale and sizeable active-sterile mixings (in this case the mass spectrum contains the three light active neutrinos, a light Majorana state with mass $M_1$ and with a  feeble mixing with the active neutrinos, and a heavy pseudo-Dirac pair with mass $M_{2,3}$). Other lepton number symmetric-inspired constructions of the type-I seesaw are possible, cf. e.g.~\cite{Shaposhnikov:2006nn}.

In some variants of the type-I seesaw realised at low scale, other singlet (from SM gauge interactions) neutrinos ($\nu_S$) are considered, as in the case for the  $\nu$-MSM~\cite{Asaka:2005pn,Asaka:2005an}, the Inverse~\cite{Schechter:1980gr, Gronau:1984ct, Mohapatra:1986bd} (ISS) and Linear~\cite{Barr:2003nn, Malinsky:2005bi}  (LSS) seesaw mechanisms;  these variants allow to have large neutrino Yukawa couplings with a comparatively low seesaw scale. We will thus focus on  mechanisms that rely on an approximate lepton number symmetry rather than on a hierarchy of scales to explain the light active neutrino masses. 
Beside their phenomenological impact, these simple extensions of the SM  may also accommodate the observed DM relic abundance, as in most successful realisations,
the lightest sterile state can account for the DM, while the other heavier states are employed to generate the masses of the active neutrinos at tree-level.

In a minimal setup\footnote{Minimal in the sense that the number of states introduced is the minimal one to explain oscillation data.}, we introduce two sets of RH neutrinos, $N_R$ and $\nu_S$, respectively, which only differ in that they have opposite lepton number assignments, $L=1$ and $L=-1$. The most general Lagrangian (regarding mass and interaction with the sterile  neutrinos) which respects lepton number conservation is then
\begin{equation}
    \mathcal{L}_{LC}=-\bar{L}_L Y_{\nu}\tilde{\Phi}N_R-\bar{\nu}_S^c M N_R + h.c.,
\end{equation}
where $Y_{\nu}$ and $M$ are Yukawa and Dirac mass matrices, respectively. After SSB the mass matrix can be arranged by blocks as\footnote{We organise the fields by blocks as $n_L\equiv\begin{pmatrix} \nu_L,\, N_R^c,\, \nu_S^c\end{pmatrix}^T$.}
\begin{equation}
    \mathcal{M}_0=\begin{pmatrix}
        0 & m_D & 0\\
        m_D^T & 0 & M\\
        0 & M^T & 0
    \end{pmatrix}.
\end{equation}
At this level, light neutrino masses are exactly zero at all orders, thanks to the lepton number symmetry, while the heavy states form Dirac pairs with masses of order $m_{N}\sim \mathcal{O}\left(\sqrt{M^2+m_D^2}\right)$. The active-heavy neutrino mixing, $\theta=m_D^* M^{-1}$, is not necessarily suppressed (considering $M$ at a scale between tens and hundreds of GeV). Only once small lepton number violating terms are introduced in the Lagrangian, light neutrinos will acquire a mass. The possible LNV terms that can be considered to complete $\mathcal{L}_{LC}$ are given below (all these terms break lepton number in two units, $\Delta L=2$)
\begin{equation}
    \mathcal{L}=\mathcal{L}_{LC}-\bar{L}_L\epsilon Y_{\nu}^{\prime}\tilde{\Phi}\nu_S-\frac{1}{2}\bar{\nu}_S\mu \nu_S^c-\frac{1}{2}\bar{N}_R^c\mu_R N_R+h.c.,
\end{equation}
which translate into the following neutrino mass matrix:
\begin{equation}
    \begin{gathered}
    \mathcal{M}_{\nu}=\mathcal{M}_0+\Delta \mathcal{M}_{LSS}+\Delta\mathcal{M}_{ISS},\,\textrm{with}\\
        \Delta \mathcal{M}_{LSS}\equiv 
        \begin{pmatrix}
            0 & 0 & \epsilon m_D^{\prime}\\
            0 & 0 & 0\\
            \epsilon\left(m_D^{\prime}\right)^T & 0 & 0
        \end{pmatrix} \, \textrm{and  }\, 
        \Delta \mathcal{M}_{ISS}\equiv 
        \begin{pmatrix}
            0 & 0 & 0\\
            0 & \mu_R & 0\\
            0 & 0 & \mu
        \end{pmatrix}.
    \end{gathered}
    \label{eq:nu_mass_matrix}
\end{equation}
Under the approximate lepton number symmetry, such that $\mu,\,\mu_R,\,\epsilon m_D^{\prime}\equiv\epsilon v_H Y_{\nu}^{\prime}/\sqrt{2}\ll m_D, M$, we can safely neglect the effect of $\mu_R$, which contributes to light neutrino masses at loop level\footnote{In the original formulation of the inverse seesaw~\cite{Mohapatra:1986bd}, the smallness of the LNV parameter, $\mu$, is linked to the supersymmetry breaking effects of an E6 scenario. In the context of $SO(10)$, which contains remnants of a larger E6 group, $\mu$ is generated at two-loop while
$\mu_R$ is generated at higher loop, justifying why we can safely neglect here the parameter $\mu_R$.}. Again, while in principle there is no reason for lepton number to be approximately conserved, the inclusion of these small LNV terms is still natural in the 't Hooft sense, as lepton number protects them from receiving large loop corrections. Depending on which sources of LNV are included in the Lagrangian, one can distinguish among different scenarios, namely the linear seesaw (when only $\Delta \mathcal{M}_{LSS}$ is included), the inverse seesaw (including only the small Majorana mass, $\mu$, for $\nu_S$), or, what we name the LISS scenario, when both $\mu$ and $\epsilon m_D^{\prime}$ contributions are included.

Upon the inclusion of these LNV terms, light neutrinos acquire masses proportional to the sources of LNV, while heavy neutrinos now form pseudo-Dirac pairs whose mass splitting is also proportional to the LNV terms. In the case of the LSS one finds for light neutrino masses
\begin{equation}
    m_{\nu}^{LSS}\sim \epsilon \left(m_D^TM^{-1}m_D^{\prime}+(m_D^{\prime})^T M^{-1}m_D\right),
\end{equation}
while for the ISS case, we have
\begin{equation}
\begin{split}
    m_{\nu}^{ISS}\sim &m_D^*\left(M^{-1}\right)^* \mu \left(M^{-1}\right)^{\dagger}m_D^{\dagger},\\
    &\Delta m_{heavy}\sim \mu,
\end{split}
\end{equation}
where we have defined the mass splitting of the pseudo-Dirac pairs as $\Delta m_{heavy}$. 

In the context of the ISS, it was found that at least 2 pairs of $N_R$ and $\nu_S$ were necessary to explain oscillation data~\cite{Abada:2014vea}, while the addition of an extra $\nu_S$ field to the particle content would additionally provide a good DM candidate at the scale $\mathcal{O}(\mu)$. The resulting model was hence labelled as ISS(2,3)~\cite{Abada:2014zra}. Indeed, for heavy states at $M\sim$~TeV, $\mu$ has to lie at the keV-scale to explain light neutrino masses, and thus the decoupled $\nu_S$ field becomes a good warm DM candidate in the few-keV range whose interactions with the SM are suppressed by powers of $\mathcal{O}\left(m_{\nu}/\mu,\mu/M\right)$ with respect to those of the SM neutrinos. On the other hand, within a type-I seesaw construction, only two $N_R$ are necessary to explain neutrino oscillations, while a third, lighter $N_R$  with suppressed Yukawa couplings can be added and identified 
as the DM candidate. %with the DM.

In the following, we will start by considering the ``single family'' case, in which we study only one SM neutrino, the DM candidate, and one Dirac pair of heavy neutrinos. This will already capture some of the most important features we will consider for the DM production, namely the existence of a heavy neutrino species capable of decaying into the SM bosons and  the DM neutrino, as well as the relevant gauge interactions with the plasma of each neutrino species. This extends over previous studies where only the DM candidate and a single SM neutrino were considered~\cite{Lello:2016rvl,Datta:2021elq}, opening the window to new production channels for the DM. Later on, we will consider full models that completely account for oscillation data as well as including the keV neutrino DM candidate, and point out regimes that cannot be described in the single family approximation.

\subsection{Minimal framework}\label{sec:minimal_frame}
As previously discussed, the main ingredients of the framework under scrutiny are represented by the presence of at least one heavy decaying neutrino, $n_h$, capable of existing in thermal equilibrium in the early stages of the history of the Universe, as well as a sterile neutrino DM candidate, $n_{DM}$. In order to ensure its cosmological stability, its coupling with the SM states should be suppressed, which motivates the assumption that it never existed in thermal equilibrium in the early Universe and its abundance is negligible prior to the freeze-in process. 

In the following, although we develop the formalism to compute and understand DM production from decays on full generality, accounting for any possible neutrino spectra and mixing between the different species, we will focus on two possibilities, motivated by Refs.~\cite{Abada:2014zra, Lucente:2021har}. For very low-scale seesaw realisations (such as the so called $\nu$-MSM, where the new states lie at the GeV scale), one generally expects production rates from decays to be suppressed because of the large mass difference between the parent particle and the Higgs boson, making the decay possible only for particles at the very end of the thermal distribution\footnote{Note however that the $\nu$-MSM - which is not our study case -  can nevertheless ensure viable DM abundance via the Shi-Fuller mechanism.}, and thus we will consider the heavy neutrinos to lie around the EW scale.

We will first consider a simplified scenario motivated by the ISS(2,3) model proposed and studied in Ref.~\cite{Abada:2014zra}, which in the single family limit presents a simplified neutrino spectrum with just one light neutrino, $n_l$, the keV DM, $n_{DM}$, and two pseudo-Dirac heavy neutrinos, $n_{h}$ with $h=1,2$. Neglecting contributions from $\mu_R$ in Eq.~(\ref{eq:nu_mass_matrix}), the mass matrix in the basis $(\nu_L,\, N_R^c,\, \nu_S^c,\, \nu_{DM}^c)^T$  is given by
\begin{equation}
    \mathcal{M}=\begin{pmatrix}
        0 & m_D & 0 & 0\\
        m_D & 0 & M & 0\\
        0 & M & \mu_{SS} & \mu_{SS^{\prime}}\\
        0 & 0 & \mu_{SS^{\prime}} & m_{DM}
    \end{pmatrix},
\end{equation}
where all elements are now just mass parameters. 
Upon diagonalisation, one finds that the active-DM mixing, $\mathcal{U}_{\alpha 4}$, and the active-heavy mixing, $\mathcal{U}_{\alpha 5(6)}$ are given by
\begin{equation}
    \mathcal{U}_{\alpha 4}\sim \theta^2 \frac{\mu_{SS^{\prime}}}{m_{DM}},\quad \mathcal{U}_{\alpha 5(6)}=\theta = \frac{m_D}{M}.
\end{equation}
From here it is clear that, in order to comply with the strong X-ray bounds on the active-DM mixing, a hierarchy must always be present either between $m_D$ and $M$ to make $\theta\ll 1$, between $\mu_{SS^{\prime}}$ and $m_{DM}$, or both. This in turn would translate into a suppression of DM production rates. Finally, the mass gap between the pseudo-Dirac pair would be of order $\mu_{SS}$. Note that this suppression of the couplings are generally to be expected in any neutrino mass model when working in the single family approximation.

Second, we will work and compare with full neutrino mass models accounting for oscillation data. In particular we will work with the type-I seesaw, where only two RH neutrinos need to be introduced in the minimal case to explain oscillations, on top of the DM candidate. This in turn translates into having one massless active neutrino. To consistently explain oscillation data when studying the parameter space, we adopt the Casas-Ibarra parametrisation introduced previously in Eq.~(\ref{eq:CI}), working in the approximate lepton number symmetric limit. In contrast to the single family approximation, it is possible to have large couplings while complying with experimental and astrophysical bounds thanks to cancellations in the matrix equations.

Both in the single family approximation as well as with complete frameworks, we will in general be able to write the relation between the flavour and mass eigenstates as
\begin{equation}
\begin{pmatrix}
\nu_L\\
N_R^c
\end{pmatrix}
= \mathcal{U}P_L
\begin{pmatrix}
    n_l\\
    n_{DM}\\
    n_h
\end{pmatrix},
\end{equation}
where $\mathcal{U}$ is the total neutrino mixing matrix whose dimensionality depends on the number of states. We have separated the mass eigenstates by blocks, with light neutrinos, $n_l$, the DM candidate with mass $m_{DM}$, $n_{DM}$, and the HNL, $n_h$, whose masses are $m_N$\footnote{Note that in the approximate lepton number conserving limit, heavy neutrinos become almost degenerate.}. The ISS can be readily accommodated by changing $N_R^c\rightarrow \begin{pmatrix} N_R^c & \nu_S^c\end{pmatrix}^T$.

\subsection{DM production}
The most popular option for the production of keV sterile neutrino DM is represented by oscillation processes in which active neutrinos are converted into DM in the early Universe. We recall that this kind of production mechanism, in its minimal incarnation, is the Dodelson-Widrow mechanism~\cite{Dodelson:1993je}. DM production via oscillation gets enhanced when occurring in an environment with a sizable lepton asymmetry, case known as the Shi-Fuller (SF) production mechanism~\cite{Shi:1998km}. In both scenarios, most of the DM is produced at relatively low temperatures, close to the QCD phase transition. Although thermal corrections are relevant also for these mechanisms, the computations discussed in this work do not strictly apply since they are aimed for much higher temperature regimes. Furthermore, the DW and SF mechanisms are nowadays very strongly constrained by both DM Indirect Detection (see e.g.~\cite{Boyarsky:2012rt} for a review) and structure formation~\cite{Schneider:2016uqi,Murgia:2017lwo}. Nonetheless, in the following, even if the DW mechanism cannot account for all the  observed DM, it will inevitably contribute to the final abundance, as it represents an unavoidable source of DM at lower temperatures. On the other hand, we will not take into account the SF production as it relies on a lepton asymmetry, which we will assume to be negligible here.

Our reference scenario will be, instead, represented by the case of DM production at earlier stages of the history of the Universe (i.e. at higher temperature regimes) via decays of heavy SM states (like the W/Z bosons) or BSM states, namely heavy sterile neutrinos. A very natural option, along this line, was proposed in~\cite{Abada:2014zra}. There, keV scale DM was embedded into the ISS(2,3) mechanism for the generation of the SM neutrino masses. This framework allows for the existence of $\mathcal{O}(0.1-1)$~TeV heavy neutrinos with sufficiently large Yukawa couplings to be in thermal equilibrium in the early Universe. These HNL can efficiently produce DM \`a-la freeze-in via their decay into a SM Higgs and a sterile neutrino. The DM relic density can be estimated as~\cite{Hall:2009bx,Chu:2011be,Chu:2013jja,Klasen:2013ypa,Blennow:2013jba,Yaguna:2011qn}
\begin{equation}\label{eq:prod-naive}
\Omega_{\rm DM}h^2 \sim 3 \times 10^{24} \left(\frac{m_{\rm DM}}{m_N^2}\Gamma ( n_{\rm h} \rightarrow n_{\rm DM}+\text{SM})+\sum_{B}\frac{m_{\rm DM}}{M_B^2}\Gamma ( B \rightarrow n_{\rm DM}+\text{SM})\right), 
\end{equation}
where an eventual sum over multiple heavy neutrino states is implicitly assumed in the first term while the second term represents the sum for decaying SM bosons, $B=H,\,W,\,\text{and}\,Z$, into DM. 
A back of the envelope estimation from Eq.~(\ref{eq:prod-naive}) leads to the fact that, neglecting production through SM boson decays, for a DM mass $m_{\rm DM}\simeq 10$~keV (typical scale for the ISS(2,3)) and  $m_N \simeq 150$~GeV for the HNL mass, a decay rate $\Gamma ( n_{\rm h} \rightarrow n_{\rm DM}+\text{SM}) \sim \mathcal{O}(10^{-16})$~GeV is needed to generate the correct amount of DM. 

In the scenario considered in Ref.~\cite{Abada:2014zra}, the DM production process considered was the decay of heavy pseudo-Dirac neutrinos into the DM itself and a Higgs boson, finding that one could indeed simultaneously explain the DM relic density with a keV sterile neutrino and the neutrino oscillation phenomenon. This conclusion was then further studied in the context of the type-I seesaw in Ref.~\cite{Lucente:2021har}, while the production through weak-gauge boson decays into DM was investigated in Ref.~\cite{Datta:2021elq}, however neglecting possible thermal effects that arise at temperatures around the EW scale~\cite{Lello:2016rvl}. 

Other processes, relevant at higher temperatures, would be for example $2\rightarrow2$ scatterings and the so-called LPM effect. The former arises, for example, through scatterings of gauge bosons with leptons producing a Higgs and the DM candidate in the final state. The rate can be estimated, at high temperatures, as $\sigma\sim \mathcal{O}\left(g^2(Y_{\nu}^{\dagger}Y_{\nu})_{ii}T^2\right)$, with $i$ the matrix element corresponding to the DM candidate. The latter arises from the resummation of multiple soft scatterings which become relevant (and can even be the dominant effect) at temperatures above the EW symmetry breaking. Thus, we expect its effect to be mild on the temperature ranges under consideration. Nonetheless, we stress that a proper inclusion of these effects is necessary in order to study the complete DM production at temperatures above $T_{SSB}\sim160$~GeV.

\subsection{Decay channels for DM production}\label{sec:cases}
The spectrum arising in the neutrino sector from simultaneously having the dark matter candidate and explaining oscillation data through heavier states, here assumed to lie around the EW scale, translates into a rich array of possibilities for DM production. 

First, the mixing between active neutrinos and DM allows for the production through weak boson decays into a DM candidate and a lepton (either a charged lepton for $W$ decays or a mostly active neutrino for $Z$ decays\footnote{Note that $Z$ decays into two DM particles are also possible, but these are suppressed by an extra power of the mixing and will thus be subdominant. Nonetheless, the formalism developed here also takes into account this negligible contribution.}). This contribution was studied in Ref.~\cite{Datta:2021elq} neglecting thermal effects, while the inclusion of thermal effects was first done in Ref.~\cite{Lello:2016rvl} in the context of Dirac neutrinos. In the following we will assume that neutrinos are Majorana particles, which quantitatively change the picture. Additionally, one could consider the Higgs boson decays into DM and a light neutrino, but given their rather light masses, this contribution is negligible as the coupling goes through the mass.

When heavy neutrinos, $n_h$, have large enough mixings with the active ones, they can be present in the plasma at electroweak temperatures, opening up the possibility for various new decay channels not present when considering only the DM candidate. Assuming they are heavier than the weak gauge bosons and the Higgs, DM can be produced through the decay of $n_h$ into a $Z$ boson and DM. Moreover, the rather large mixings necessary for the heavy neutrinos to be in thermal equilibrium translate into non-negligible couplings with the Higgs boson, making the production channel $n_h\rightarrow H+n_{DM}$ relevant. What is more, from the statistical distribution of the final states, the decay into the Higgs boson could potentially show an enhancement when compared to the decay of a boson into a couple of fermions. This is thanks to the fact that the decay to a scalar can have Bose enhancement, while a pair of fermions in the final state tends to suppress the rate with respect to the zero temperature case due to Pauli blocking~\cite{Lundberg:2020mwu}. 

Given the richness in the possible decays to DM, in the following we will study several scenarios which introduce successively a new contribution to the production rate. We thus consider the following cases, that are summarized in Table~\ref{tab:summary_scenarios} with their production channels:
\begin{itemize}
    \item Considering solely one species of active neutrinos and the DM candidate, without the introduction of HNL. This is the simplest scenario, which would correspond as well to the case studied originally in the proposal of the Dodelson-Widrow mechanism~\cite{Dodelson:1993je}. In this case the only relevant parameters are the active-DM mixing and the DM mass. For the production through decays, only weak-gauge bosons could produce a non-negligible amount of DM, given that the Higgs couples to the rather light masses. Although previously studied in Ref.~\cite{Lello:2016rvl} in the context of Dirac neutrinos, here we will instead consider them to be Majorana. When referring to this simple case, we will call it the ``toy $2\times 2$'' scenario, given that it could describe DM but not account for %totally neglects 
    neutrino masses and mixings.

    \item Introducing HNL in an ISS-like construction within the single family approximation. This is inspired by Ref.~\cite{Abada:2014zra} where it was found that the ISS(2,3) could simultaneously explain oscillation data and the DM abundance through a keV-scale candidate. In our simplified one-family ISS, we will thus have one light neutrino, the DM candidate, and a pair of almost degenerate HNL, and thus refer to it as ``ISS-$4\times 4$''. We choose the parameters to comply with current constraints on mixing between the different species both for the DM and the heavy species, but point out that this case does not explicitly reproduce oscillation data, although there is in principle enough freedom when going to the full case to fit data. In any case it proves useful because it already introduces the non-negligible coupling between the Higgs boson and the heavy neutrinos, as well as the possible decay of heavy neutrinos into a gauge boson and the DM neutrino.

    \item Minimal type-I seesaw, fully reproducing oscillation data through the Casas-Ibarra parametrisation~\cite{Casas:2004gh} from Eq.~(\ref{eq:CI}), with two almost degenerate HNL and the DM candidate. We choose the type-I as the coupling between DM, HNL and the Higgs is parametrically different in terms of Yukawa couplings to the one of the ISS. Additionally, we will first neglect the Higgs contribution to fully study how the addition of HNL changes the production compared to the ``toy $2\times2$'' scenario, and finally introduce also the Higgs contribution, which is non-negligible, and compare with the results from the ``ISS-$4\times 4$''.
    
\end{itemize}

\begin{table}[h]
    \centering
    \begin{tabular}{|c||c|}
    \hline
    Scenario & Decay channels \\
    \hline
    \hline
    Toy~$2\times2$ & $W(Z)\rightarrow \ell_{\alpha} (n_{l}) +n_{DM}$\\
    \hline
    ISS~$4\times4$ & $W(Z)\rightarrow \ell_{\alpha} (n_{l}) +n_{DM}$, $H\rightarrow n_{l}+n_{DM}$, $n_h\rightarrow H(Z) + n_{DM}$\\
    \hline
    Type-I seesaw w/o Higgs & $W(Z)\rightarrow \ell_{\alpha} (n_{l}) +n_{DM}$, $n_h\rightarrow Z + n_{DM}$\\
    \hline
    Type-I seesaw & $W(Z)\rightarrow \ell_{\alpha} (n_{l}) +n_{DM}$, $H\rightarrow n_{l}+n_{DM}$, $n_h\rightarrow H(Z) + n_{DM}$\\
    \hline
    \end{tabular}
    \caption{Summary of the possible DM production channels through two-body decays available in the different cases we consider. Note that we assume here that the HNL ($n_h$) are heavier than the Higgs. In the opposite case, one could still have the decay $H\rightarrow n_h + n_{DM}$. The difference between the ``ISS~$4\times4$'' and the full type-I seesaw is that in the latter the matrix structure of the couplings allows for regimes in which cancellations between elements can allow for large couplings.}
    \label{tab:summary_scenarios}
\end{table}

\section{Production rates in thermal field theory}\label{Sec:self-energy}
\subsection{Formalism and strategy}
We will make use of the \textit{real-time formalism} of thermal QFT~\cite{Schwinger:1960qe,Keldysh:1964ud} in order to compute decay rates and dispersion relations for the collective propagating modes in the plasma~\cite{le_bellac_1996,kapusta_gale_2006,Lundberg:2020mwu}. Our aim is to generalise the results of Ref.~\cite{Lello:2016rvl} to any neutrino mass model which includes SM-singlet fermions, without any assumption on the size of the mixing matrix elements, $\mathcal{U}_{\alpha i}$, or the neutrino masses. This will allow to understand whether the presence of heavier neutrinos mixing with the DM can change the picture described in Ref.~\cite{Lello:2016rvl} where only a light neutrino and the DM were considered.

In the \textit{real-time formalism} we have a doubling of the degrees of freedom, such as the free propagator becomes a $2\times 2$ matrix. For a particle of mass $m$, taking the symmetric-Keldysh contour and in the absence of a chemical potential,  
its components can be written as~\cite{le_bellac_1996}
\begin{equation}
\begin{cases}
    \tilde{D}^{(++)}_{\alpha\beta}(k)=\left[\frac{i}{k^2-m^2+i\epsilon}+2\pi\eta f(|k_0|) \delta(k^2-m^2)\right]d_{\alpha\beta},\\
    \tilde{D}_{\alpha\beta}^{(--)}(k)=[\tilde{D}^{(++)}(k)]^*d_{\alpha\beta},\\
    \tilde{D}_{\alpha\beta}^{(+-)}(k)=e^{\beta k_0/2}f(k_0)2\pi\varepsilon(k_0)\delta(k^2-m^2)d_{\alpha\beta},\\
    \tilde{D}_{\alpha\beta}^{(-+)}(k)=\eta \tilde{D}_{\alpha\beta}^{(+-)}(k),
\end{cases}
\label{eq:propagators2x2}
\end{equation}
where $f(k_0)$ is the equilibrium distribution function given by
\begin{equation}
    f(k_0)=\frac{1}{e^{\beta k_0}-\eta},
\end{equation}
with $\eta=\pm 1$ for bosons and fermions, respectively, and $\beta$ is the inverse of the temperature of the plasma. The factor $d_{\alpha\beta}$ encodes all the Lorentz structure of the propagator, and is given by
\begin{equation}
d_{\alpha\beta}=
    \begin{cases}
    1, & \text{for scalars},\\
    \slashed{k}\pm m\mathbb{1}, & \text{for fermions},\\
    -\eta_{\mu\nu}+\frac{k_{\mu}k_{\nu}}{m^2}, & \text{for vector bosons}.
    \end{cases}
\end{equation}
Even if one has to deal with a matrix structure for the propagator, there are some relations between the components of the propagators which can help simplify computations. In particular, the following relation can be found between the retarded 
self-energy of a fermion, $\sigma^R$, and the $(+-)$-component of the self-energy in the \textit{real-time formalism}~\cite{le_bellac_1996}: 
\begin{equation}
    \mathrm{Im} \sigma^R = -\cosh{\left(\frac{\beta |p_0|}{2}\right)}\left[\Theta(p_0)-\Theta(-p_0)\right]i \sigma^{(+-)}\ .
    \label{eq:Ret_Im_sigma}
\end{equation}
It is indeed the imaginary part of the retarded 
self-energy that will eventually be related  to the dumping rates of quasi-particle excitations in the plasma~\cite{le_bellac_1996}. Regarding the real part of $\sigma^R$, it can be obtained through the following dispersion relation~\cite{Lello:2016rvl}
\begin{equation}
    \mathrm{Re} \sigma^R = \frac{1}{\pi}\int_{-\infty}^{\infty} d q_0 \mathcal{P}\left[\frac{\mathrm{Im}\sigma^R (q_0)}{p_0-q_0}\right].
    \label{eq:dispersion_relation}
\end{equation}
As will be apparent in the next section, in general we will be able to write the self-energy contributions for massive neutrinos as
\begin{equation}\label{eq:Sigma}
    \Sigma_{ij}\propto \sum_k \mathcal{A}_{ik}\mathcal{A}_{kj}\sigma^R_k,
\end{equation}
where $\mathcal{A}_{ij}$ are some combination of coupling constants appearing at the Lagrangian level, and the index $k$ runs over mass eigenstates. Thus, we can still make use of the dispersion relation for $\sigma_k$ in order to find the real and imaginary parts of $\Sigma$. Given that the couplings $\mathcal{A}_{ij}$ do not depend on momenta, in the following we can %just 
 use the relation from Eq.~(\ref{eq:Ret_Im_sigma}) including them in Eq.~(\ref{eq:Sigma}). When $\mathcal{A}_{ij}\in\mathbb{R}$, this quantity will be directly $\mathrm{Im}\Sigma_{ij}$, but this will not necessarily be the case. Thus, in the following we will dub the product $\mathcal{A}_{ik}\mathcal{A}_{kj}\mathrm{Im}\sigma_k^R$ as $\mathcal{I}\left(\Sigma_{ij}\right)$ and $\mathcal{A}_{ik}\mathcal{A}_{kj}\mathrm{Re}\sigma_k^R$ will be $\mathcal{R}\left(\Sigma_{ij}\right)$. It is easy to note that $\mathcal{R}\left(\Sigma_{ij}\right)+i\mathcal{I}\left(\Sigma_{ij}\right)=\mathrm{Re}\Sigma_{ij}+i \mathrm{Im}\Sigma_{ij}$, which is the final quantity we are interested in.

On general grounds, in the following, we will %be able to 
decompose the self-energy as (suppressing generation indices to ease the notation)
\begin{equation}
    \Sigma = \gamma_0 \Sigma^{(0)}-\vec{\gamma}\cdot \hat{p}\Sigma^{(1)}+\Sigma^{(2)},
\end{equation}
with $\hat{p}$ an unit vector in the direction of the momentum $\vec{p}$. Thanks to this decomposition, we will be able to project $\Sigma$ and work with the scalar functions $\Sigma^{(I)}$ instead, obtained as 
\begin{equation}
\begin{gathered}
    \Sigma^{(0)}=\frac{1}{4}\mathrm{Tr}\left[\gamma^0 \Sigma\right],\,
    \Sigma^{(1)}=\frac{1}{4}\mathrm{Tr}\left[\hat{p}\cdot \vec{\gamma}\Sigma\right],\,\textrm{and}\,
    \Sigma^{(2)}=\frac{1}{4}\mathrm{Tr}\left[\Sigma\right].
\end{gathered}
\label{eq:decomp_Sigma}
\end{equation}
When necessary, we will also take into account the chirality projectors $P_{L(R)}$. As shown in Ref.~\cite{Weldon:1983jn}, once the one-loop self-energy corrections are computed, the rate at which a species in the plasma reaches equilibrium will be given by the imaginary part of the self-energy. This is precisely the quantity we are interested in within the freeze-in framework. The advantage of using the one-loop corrections is that it directly includes all possible interactions with the plasma, as well as the correct dependence with temperature and momenta for the production rates.

\subsection{Higgs contribution}\label{sec:H_cont}
\begin{figure}
\centering
\begin{tikzpicture}
  \begin{feynman}
    \vertex (a) {$n_i$};
    \vertex [right=1cm of a] (b);
    \vertex [right=3cm of b] (c);
    \vertex [right=1cm of c] (f2) {$n_j$};

    \diagram[horizontal=a to b]{
      (a) -- [plain] (b),
      (b) -- [plain, edge label=$n_k$, momentum'=$q$] (c),
      (b) -- [scalar, edge label=$H$, half left, momentum'=$p-q$] (c),
      (c) -- [plain] (f2),
    };
  \end{feynman}
\end{tikzpicture}
\begin{tikzpicture}
    \begin{feynman}
    \vertex (a) {$n_i$};
    \vertex [right=1cm of a] (b);
    \vertex [right=3cm of b] (c);
    \vertex [right=1cm of c] (f2) {$n_j$};

    \diagram[horizontal=a to b]{
      (a) -- [plain] (b),
      (b) -- [plain, edge label=$n_k\,(\ell_{\alpha})$, momentum'=$q$] (c), % The figure does not compile if there is a comme in label, e.g. label=$n_k \ell$. I don't know why
      (b) -- [boson, edge label=${Z\,(W)}$, half left, momentum'=$p-q$] (c),
      (c) -- [plain] (f2),
    };
  \end{feynman}
\end{tikzpicture}
\caption{Feynman diagrams contributing to the neutrino self-energy. In the SM extension with singlet fermions, there is the Higgs contribution (left) an\textbf{d} the gauge ($W$ and $Z$ bosons) contribution from the $W$ and $Z$ bosons (right). The imaginary part of the self-energy will be related to the rate at which each species would reach equilibrium.}
\label{fig:self-energies}
\end{figure}
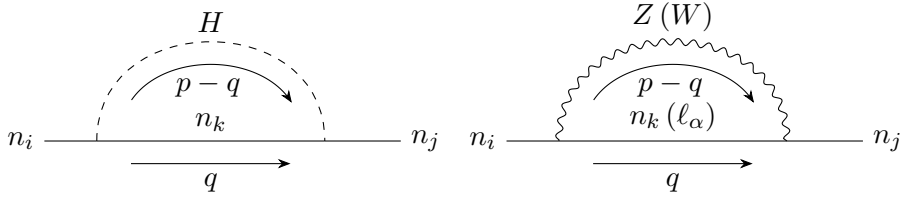

We study here the thermal self-energy contribution from the Higgs boson coupling, which after SSB can be written as~\cite{Alonso:2012ji}
\begin{equation}
\begin{split}
    \mathcal{L}_{H-\nu}\supset& -\frac{H}{v_H}\sum_{i,j}C_{ij}\bar{n}_i\left(m_i P_L+m_j P_R\right)n_j=\\
    &-\frac{H}{2v_H}\sum_{i,j}\bar{n}_i\left[C_{ij}\left(m_i P_L+m_j P_R\right)+C^*_{ij}\left(m_i P_R+m_j P_L\right)\right]n_j,
    \label{eq:lag_Higgs-nu}
\end{split}
\end{equation} where $C_{ij}\equiv \sum_{\alpha=e,\mu,\tau}\mathcal{U}^{\dagger}_{i \alpha}\mathcal{U}_{\alpha j}$ and $m_{i(j)}$ are the different neutrino masses and the second line uses the fact that neutrinos are Majorana. % ($v_H$ is the Higgs vaccum expectation value).
Using the propagators from Eq.~(\ref{eq:propagators2x2}), we obtain
\begin{equation}
\begin{split}
    i\Sigma_{ij}^{(+-)}=&-\frac{2\pi}{v_H^2}e^{\frac{\beta p_0}{2}}f_F(p_0)\sum_k \int\frac{d^4q}{(2\pi)^3}\left[1+f_B(r_0)-f_F(q_0)\right]\varepsilon(q_0)\varepsilon(r_0)\\ &\left[\left(\mathcal{A}_{kj}^R\slashed{q}+\mathcal{A}_{kj}^Lm_k\right)\mathcal{A}_{ik}^LP_L+\left(\mathcal{A}_{kj}^L\slashed{q}+\mathcal{A}_{kj}^Rm_k\right)\mathcal{A}_{ik}^RP_R\right]\delta(q^2-m_k^2)\delta(r^2-M_H^2),
\end{split}
\label{eq:pm_selfenergy}
\end{equation}
where $r_{\mu}\equiv (p-q)_{\mu}$ and $f_{B(F)}$ corresponds to the distribution function for bosons (fermions). To ease the notation, we have defined $\mathcal{A}_{ij}^L\equiv \frac{1}{2}\left(C_{ij} m_i+C_{ij}^*m_j\right)$ and $\mathcal{A}_{ij}^R=\left(\mathcal{A}_{ij}^L\right)^*$. From this expression we readily obtain
\begin{equation}
\begin{split}
    \mathcal{I}\left(\Sigma_{ij}\right)=&\frac{\pi}{v_H^2}\sum_{k}\int \frac{d^4q}{(2\pi)^3}\left[1+f_B(r_0)-f_F(q_0)\right]\varepsilon(q_0)\varepsilon(r_0)\\ &\left[\left(\mathcal{A}_{kj}^R\slashed{q}+\mathcal{A}_{kj}^Lm_k\right)\mathcal{A}_{ik}^LP_L+\left(\mathcal{A}_{kj}^L\slashed{q}+\mathcal{A}_{kj}^Rm_k\right)\mathcal{A}_{ik}^RP_R\right]\delta(q^2-m_k^2)\delta(r^2-M_H^2),
\end{split}
\end{equation}
which can be furthered simplified by decomposing the Dirac delta functions as
\begin{equation}
\delta(q^2-m_k^2)=\frac{1}{2\omega_k}\left[\delta(q_0-\omega_k)+\delta(q_0+\omega_k)\right],
\end{equation}
with\footnote{Similarly, for $\delta(r^2-M_H^2)$ we would have $\omega_H=\sqrt{(\vec{p}-\vec{q})^2+M_H^2}$.} $\omega_k\equiv \sqrt{\vec{q}^2-m_k^2}$, leading to
\begin{equation}
\begin{split}
    \mathcal{I}\left(\Sigma_{ij}\right)=&\frac{\pi}{v_H^2}\varepsilon(p_0)\sum_{k}\int_{-\infty}^{\infty}dq_0\int \frac{d^3q}{(2\pi)^3}\frac{1}{4\omega_k\omega_H}\left[1+f_B(r_0)-f_F(q_0)\right]\\ &\left[\left(\mathcal{A}_{kj}^R\slashed{q}+\mathcal{A}_{kj}^Lm_k\right)\mathcal{A}_{ik}^LP_L+\left(\mathcal{A}_{kj}^L\slashed{q}+\mathcal{A}_{kj}^Rm_k\right)\mathcal{A}_{ik}^RP_R\right]\\
    &\left[\delta(q_0-\omega_k)-\delta(q_0+\omega_k)\right]\left[\delta(r_0-\omega_H)-\delta(r_0+\omega_H)\right].
\end{split}
\label{eq:ImSigma_1}
\end{equation}
Notice that the latter expression agrees with Ref.~\cite{Wu:2009yr} in the limit $m_{i(j)}\rightarrow 0$.

Projecting the self-energy matrix as prescribed in Eq.~(\ref{eq:decomp_Sigma}) we obtain
\begin{equation}
\begin{split}
    \mathcal{I}\left(\Sigma_{ij}^{(I)}\right)\bigg|_{L(R)}=&\frac{\pi}{v_H^2}\varepsilon(p_0)\sum_{k}\int_{-\infty}^{\infty}dq_0\int \frac{d^3q}{(2\pi)^3}\frac{\xi^{(I)}_{L(R)}}{4\omega_k\omega_H}\left[1+f_B(r_0)-f_F(q_0)\right]\\
    &\left[\delta(q_0-\omega_k)-\delta(q_0+\omega_k)\right]\left[\delta(r_0-\omega_H)-\delta(r_0+\omega_H)\right],
\end{split}
\end{equation}
where the sub-index $L(R)$ corresponds to the different chiralities and the ``prefactors'' $\xi^{(I)}_{L(R)}$ are 
\begin{equation}
\xi^{(I)}_L=
    \begin{cases}
    A^R_{kj}\mathcal{A}_{ik}^L q_0, & \text{for }I=0,\\
    -A^R_{kj}\mathcal{A}_{ik}^L\frac{\vec{p}\cdot\vec{q}}{|\vec{p}|}, & \text{for }I=1,\\
    A^L_{kj}\mathcal{A}_{ik}^L, & \text{for }I=2,
    \end{cases}
\end{equation}
while one can notice that $\xi^{(I)}_{R}=\left(\xi^{(I)}_L\right)^*$. Thanks to this decomposition, we can obtain $\mathcal{I}\left(\Sigma_{ij}^{(I)}\right)$ in terms of some constant pre-factors and the following three master integrals\footnote{From now on to ease notation we write $p\equiv |\vec{p}|$.} 
\begin{equation}
    \begin{gathered}
    I^{(0)}\equiv \int \frac{d^4q}{(2\pi)^3}\frac{q_0}{4\omega_k\omega_H}\left[1+f_B(r_0)-f_F(q_0)\right]
    \left[\delta(q_0-\omega_k)-\delta(q_0+\omega_k)\right]\left[\delta(r_0-\omega_H)-\delta(r_0+\omega_H)\right],\\
    I^{(1)}\equiv\frac{1}{p}\int \frac{d^4q}{(2\pi)^3}\frac{\vec{p}\cdot \vec{q}}{4\omega_k\omega_H}\left[1+f_B(r_0)-f_F(q_0)\right]
    \left[\delta(q_0-\omega_k)-\delta(q_0+\omega_k)\right]\left[\delta(r_0-\omega_H)-\delta(r_0+\omega_H)\right],\\
    I^{(2)}\equiv \int \frac{d^4q}{(2\pi)^3}\frac{1}{4\omega_k\omega_H}\left[1+f_B(r_0)-f_F(q_0)\right]
    \left[\delta(q_0-\omega_k)-\delta(q_0+\omega_k)\right]\left[\delta(r_0-\omega_H)-\delta(r_0+\omega_H)\right].
    \end{gathered}
    \label{eq:integrals_higgs1}
\end{equation}
It will prove useful to change variables as 
\begin{equation}
\int d^3q=\int q^2 dq d\varphi d(\cos{\theta})=-\int\frac{q}{p}\omega_Hdqd\varphi d\omega_H,
\end{equation}
where the variable $\omega_H$ has limits $\omega_H^{\pm}=\sqrt{(p\pm q)^2+M_H^2}$. Upon this change of variables, the latter integrals become
\begin{equation}\label{eq:master2}
    \begin{gathered}
        I^{(0)}\equiv \frac{1}{p}\int_{-\infty}^{\infty}dq_0\int_0^\infty dq\int_{\omega_H^-}^{\omega_H^+}d\omega_H \frac{qq_0}{(2\pi)^24\omega_k}\left[1+f_B(r_0)-f_F(q_0)\right]\\
        \left[\delta(q_0-\omega_k)-\delta(q_0+\omega_k)\right]\left[\delta(r_0-\omega_H)-\delta(r_0+\omega_H)\right],\\
        I^{(1)}\equiv\frac{1}{p^2}\int_{-\infty}^{\infty}dq_0\int_0^\infty dq\int_{\omega_H^-}^{\omega_H^+}d\omega_H\frac{q}{(2\pi)^24\omega_k}\left(p_0q_0-p \tilde{\mu}\right)\left[1+f_B(r_0)-f_F(q_0)\right]\\
        \left[\delta(q_0-\omega_k)-\delta(q_0+\omega_k)\right]\left[\delta(r_0-\omega_H)-\delta(r_0+\omega_H)\right],\\
        I^{(2)}\equiv \frac{1}{p}\int_{-\infty}^{\infty}dq_0\int_0^\infty dq\int_{\omega_H^-}^{\omega_H^+}d\omega_H\frac{q}{(2\pi)^24\omega_k}\left[1+f_B(r_0)-f_F(q_0)\right]\\
        \left[\delta(q_0-\omega_k)-\delta(q_0+\omega_k)\right]\left[\delta(r_0-\omega_H)-\delta(r_0+\omega_H)\right],
    \end{gathered}
\end{equation}
where $\tilde{\mu}= \left(p_0^2+m_k^2-p^2-M_H^2\right)/(2p)$ is an energy variable introduced in order to better identify the regions of support of the $\delta$ functions in Eq.~(\ref{eq:integrals_higgs1}), as detailed in  Appendix~\ref{app:regions_support} (and not to be confused with the LNV scale introduced in Section~\ref{sec:minimal_frame}).

\subsection{Gauge (Z) contribution}
The interaction between massive neutrinos and the $Z$ boson is given by
\begin{equation}
\begin{split}
\mathcal{L}_{Z-\nu}\supset &\frac{g}{2c_W}\sum_{i,j}C_{ij}\bar{n}_i \gamma^{\mu}P_Ln_j Z_{\mu}=\frac{g}{4c_W}\sum_{i,j}\bar{n}_i\left[C_{ij}\gamma^{\mu}P_L-C^*_{ij}\gamma^{\mu}P_R\right]n_j Z_{\mu},
\end{split}
\end{equation}
with $g$ the $SU(2)_L$ coupling constant and $c_W\equiv \cos{\theta_W}$ the cosine of the weak mixing angle. Just like in the Higgs case, the second equality holds when neutrinos are Majorana. Following the same procedure as in Section~\ref{sec:H_cont},  we arrive at
\begin{equation}
\begin{split}
\mathcal{I}\left(\Sigma_{ij}\right)=&\left(\frac{g}{2c_W}\right)^2\pi\sum_{k}\int_{-\infty}^{\infty}d q_0\int\frac{d^3 q}{(2\pi)^3}\frac{1}{4\omega_k\omega_Z}\left[1+f_B(r_0)-f_F(q_0)\right]\\
&\frac{1}{4}\gamma^{\mu}\left\lbrace\left(C_{kj}\slashed{q}-C_{kj}^*m_k\right)C_{ik}\gamma^{\nu}P_L+\left(C_{kj}^*\slashed{q}-C_{kj}m_k\right)C_{ik}^*\gamma^{\nu}P_R\right\rbrace\left(-\eta_{\mu\nu}+\frac{r_{\mu}r_{\nu}}{M_Z^2}\right)\\
&\left[\delta(q_0-\omega_k)-\delta(q_0+\omega_k)\right]\left[\delta(r_0-\omega_Z)-\delta(r_0+\omega_Z)\right],
\end{split}
\label{eq:Im_SigZ_1}
\end{equation}
with $\omega_Z\equiv\sqrt{(\vec{p}-\vec{q})^2+M_Z^2}$ and $\gamma^{\mu}$ the Dirac gamma matrices. Using the relations from Eq.~(\ref{eq:decomp_Sigma}) we find
\begin{equation}
\begin{split}
\mathcal{I}\left(\Sigma_{ij}^{(I)}\right)\bigg|_{L(R)}=&\left(\frac{g}{2c_W}\right)^2\pi\sum_{k}\int_{-\infty}^{\infty}d q_0\int\frac{d^3 q}{(2\pi)^3}\frac{\mathscr{P}_{L(R)}^{(I)}}{4\omega_k\omega_Z}\left[1+f_B(r_0)-f_F(q_0)\right]\\
&\left[\delta(q_0-\omega_k)-\delta(q_0+\omega_k)\right]\left[\delta(r_0-\omega_Z)-\delta(r_0+\omega_Z)\right],
\end{split}
\end{equation}
where the ``prefactors'' $\mathscr{P}_{L(R)}^{(I)}$ are given by 
\begin{equation}
\mathscr{P}_L^{(I)}=
    \begin{cases}
    \frac{1}{4}C_{ik}C_{kj}\left[A_0(p_0) p_0+B(p_0)q_0\right], & \text{for }I=0,\\
    -\frac{1}{8p}C_{ik}C_{kj}\left[A_1(p_0)+2p_0 B(p_0)q_0\right], & \text{for }I=1,\\
    \frac{3}{2}C_{ik}C_{kj}^*m_k, & \text{for }I=2.
    \end{cases},
\end{equation}
with the functions $A_0$, $A_1$ and $B$ given below
\begin{equation}
\begin{gathered}
    A_0(p_0)\equiv \frac{p_0^2-p^2-M_Z^2-m_k^2}{M_Z^2},\\
    B(p_0)\equiv 1-A_0(p_0),\\
    A_1(p_0)\equiv p^2+M_Z^2-p_0^2-m_k^2+A_0(p_0)\left[p^2+m_k^2-M_Z^2+p_0^2\right].
\end{gathered}
\end{equation}
The prefactors $\mathscr{P}^{(I)}_R$ can be obtained by changing $C_{ij}\rightarrow C_{ij}^*$ in the expressions of $\mathscr{P}^{(I)}_L$. Thus, we can define the following master integrals for the $Z$ gauge boson case\footnote{Performing a similar change of variables as for the Higgs case we find $\omega_Z^{\pm}=\sqrt{(p\pm q)^2+M_Z^2}$.}
\begin{equation}
    \begin{gathered}
        I_Z^{(0)}\equiv \frac{1}{p}\int_{-\infty}^{\infty}dq_0\int_0^\infty dq\int_{\omega_Z^-}^{\omega_Z^+}d\omega_Z \frac{q}{(2\pi)^24\omega_k}\left[A_0(p_0)p_0+B(p_0)q_0\right]\left[1+f_B(r_0)-f_F(q_0)\right]\\
        \left[\delta(q_0-\omega_k)-\delta(q_0+\omega_k)\right]\left[\delta(r_0-\omega_Z)-\delta(r_0+\omega_Z)\right],\\
        I_Z^{(1)}\equiv\frac{1}{p^2}\int_{-\infty}^{\infty}dq_0\int_0^\infty dq\int_{\omega_Z^-}^{\omega_Z^+}d\omega_Z\frac{q}{2(2\pi)^24\omega_k}\left[A_1(p_0)+2p_0 B(p_0)q_0\right]\left[1+f_B(r_0)-f_F(q_0)\right]\\
        \left[\delta(q_0-\omega_k)-\delta(q_0+\omega_k)\right]\left[\delta(r_0-\omega_Z)-\delta(r_0+\omega_Z)\right],\\
        I^{(2)}_Z=I^{(2)}\left(\omega_H\rightarrow \omega_Z\right).
    \end{gathered}
\end{equation}

\subsection{$W$-boson contribution}
The interaction for the $W$-boson when including HNL is given by
\begin{equation}
    \mathcal{L}_{W-\nu}\supset \frac{g}{\sqrt{2}}\sum_{i,\alpha}\mathcal{U}_{\alpha i}\bar{\ell}_{\alpha}\gamma^{\mu}P_Ln_iW^{-}_{\mu}+h.c.,
\end{equation}
Even if the $W$-boson also contributes to the neutrino self-energy corrections, the structure of the Lagrangian is such that these terms can be readily taken into account from the $Z$-boson ones. Indeed, by making the changes $g/(2c_W)\rightarrow g/\sqrt{2}$, $M_Z\rightarrow M_W$, $m_k\rightarrow m_\ell\sim 0$\footnote{Given that we are interested in temperatures at the EW scale, we neglect in the following charged lepton masses, $m_\ell$.} and $\sum_{k}C_{ik}C_{kj}\rightarrow 2\sum_{\alpha}\mathcal{U}_{i\alpha}^{\dagger}\mathcal{U}_{\alpha j}=2C_{ij}$, and noticing that in the $W$ case we get $\mathscr{P}_R^{(I)}=0$ and $\mathscr{P}_L^{(2)}=0$ identically, one can obtain the neutrino self-energy corrections from $W$-boson exchange using the same master integrals as for the $Z$ case.

\subsection{Propagating states in the medium}\label{Sec:comparison}
Given the self-energy corrections to the propagator, written on full generality as\footnote{In general $\Sigma$ is a function of $T,\,p_0$, and $p$, but we are interested here in its dependence with $p_0$ and thus omit the others to ease the notation.}
\begin{equation}
\slashed{\Sigma}(p_0) = \left(\gamma_0 \Sigma_L^{(0)}-\vec{\gamma}\cdot \hat{p}\Sigma_L^{(1)}+\Sigma_L^{(2)}\right)P_L+\left(\gamma_0 \Sigma_R^{(0)}-\vec{\gamma}\cdot \hat{p}\Sigma_R^{(1)}+\Sigma_R^{(2)}\right)P_R,
\label{eq:full_self-energy}
\end{equation}
we are interested in finding the propagating states in the medium, which follow Dirac's equation
\begin{equation}
\left(\slashed{p}-\mathcal{M}+\slashed{\Sigma}(p_0)\right)\tilde{\Psi}=0.
\end{equation}
It proves useful to decompose the wave function $\tilde{\Psi}$ in terms of helicity and chirality components, as~\cite{Lello:2016rvl}
\begin{equation}
\tilde{\Psi}=\sum_{h=\pm 1}v^h\otimes
\begin{pmatrix}
\varphi^h\\
\zeta^h
\end{pmatrix},\quad \vec{\sigma}\cdot \hat{p} v^h = h v^h,
\label{eq:helicity_decomp}
\end{equation}
where the second term in Eq.~(\ref{eq:helicity_decomp}) is the helicity operator, with the possible eigenvalues $h=\pm 1$. By using this decomposition of $\tilde{\Psi}$, together with Eq.~(\ref{eq:full_self-energy}), we arrive at the following set of coupled equations
\begin{equation}
\begin{split}
\left[p_0+\Sigma^{(0)}_R-h p-h \Sigma^{(1)}_R\right]\zeta^h-\left(\mathcal{M}+\Sigma^{(2)}_L\right)\varphi^h=0,\\
\left[p_0+\Sigma^{(0)}_L+h p+h \Sigma^{(1)}_L\right]\varphi^h-\left(\mathcal{M}+\Sigma^{(2)}_R\right)\zeta^h=0.
\end{split}
\label{eq:motion_LR}
\end{equation}
From the first line in Eq.~(\ref{eq:motion_LR}) we can solve for $\zeta^h$ as
\begin{equation}
\zeta^h=\left[p_0+\Sigma^{(0)}_R-h p-h \Sigma^{(1)}_R\right]^{-1}\left(\mathcal{M}+\Sigma^{(2)}_L\right)\varphi^h,
\end{equation}
and substitute in the second line in order to find the equation of motion for $\varphi^h$ as
\begin{equation}
\begin{split}
&\bigg[p_0^2-p^2+\left(p_0-hp\right)\left(\Sigma^{(0)}_L+h\Sigma^{(1)}_L\right)\\
&-\left(p_0-h p\right)\left(\mathcal{M}+\Sigma^{(2)}_R\right)\left[p_0+\Sigma^{(0)}_R-h p-h \Sigma^{(1)}_R\right]^{-1}\left(\mathcal{M}+\Sigma^{(2)}_L\right)\bigg]\varphi^h=0.
\end{split}
\label{eq:motion_left}
\end{equation}
We can identify the terms in brackets in Eq.~(\ref{eq:motion_left}) as the perturbed inverse propagator for left-handed fields in the medium, $\mathcal{S}^{-1}$. Indeed, in the absence of self-energy corrections we recover the usual dispersion relation in vacuum:
\begin{equation}
\left(p_0^2-p^2-\mathcal{M}^2\right)\varphi^h=0\rightarrow p_0^2=p^2+\mathcal{M}^2.
\end{equation}
In the following, what we will need to do is to precisely find the dispersion relations for the neutrinos in the hot plasma, given by the complex zeroes of the inverse propagator from Eq.~(\ref{eq:motion_left}).

\subsection{Simplified 2 neutrino case}
It is instructive to consider an example where we only have the active neutrinos and the sterile one which will become the DM candidate. This allows to simplify expressions and obtain an analytical understanding of the different terms contributing to the production rates. In this ``toy $2\times 2$'' scenario, the neutrino mass matrix, in the interaction basis, can be written as
\begin{equation}
    \mathcal{M}=\begin{pmatrix}
        0 & m_D\\
        m_D & m_{DM}
    \end{pmatrix},
\end{equation}
with $\tan{2\theta}\equiv 2m_D/m_{\rm DM}\ll 1$, where $\theta$ is the active-heavy mixing angle. 

The fact that we are interested in the production at temperatures around the EW scale allows to neglect the comparably lighter neutrino masses in the self-energy corrections, greatly simplifying Eq.~(\ref{eq:motion_left}). Additionally, we can work in the original flavour basis, in which the self-energy contributions become\footnote{Compared with the typical temperature for production and the masses of the intermediate bosons in the self-energies, both the light and the DM neutrino state masses can be neglected when evaluating the self-energies.}
\begin{equation}
    \Sigma_{L(R)}^{(I)}=\begin{pmatrix}
        \sigma_{L(R)}^{(I)} & 0\\
        0 & 0
    \end{pmatrix},
\end{equation}
where $\sigma^{(I)}_{L(R)}$ is an appropriate combination of the $W$ and $Z$ boson contributions to the retarded 
self-energies\footnote{In particular, the $W$ boson does not contribute to the RH part of the self-energies as is clear from the Lagrangian interaction.}. Thus, neglecting the contributions from $\Sigma^{(2)}$, the inverse propagator, $\mathcal{S}^{-1}$, becomes
\begin{equation}
    \mathcal{S}^{-1}=\left(p_0^2-p^2\right)\mathbb{I}_{2\times 2}+\begin{pmatrix}
        \Omega^h - \frac{m_{DM}^2}{4}\tan^2{2\theta} & -\frac{m_{DM}^2}{2}\tan{2\theta}\\
        -\frac{m_{DM}^2}{2}\tan{2\theta} & -m_{DM}^2\left(1+\frac{1}{4\alpha^h}\tan^2{2\theta}\right)
    \end{pmatrix},
    \label{eq:Matrix_Inverse_Propagator}
\end{equation}
where we have defined the following combinations of self-energy corrections, that depend on the helicity eigenvalue ($h=\pm1$):
\begin{equation}
    \begin{gathered}
        \Omega^h\equiv (p_0-h p)\left(\sigma_L^{(0)}+ h\sigma_L^{(1)}\right),\\
        \alpha^h\equiv 1+(p_0-hp)^{-1}\left(\sigma_R^{(0)}-h \sigma_R^{(1)}\right).
        \label{eq:self_energy_helicity}
    \end{gathered}
\end{equation}
In order to find the dispersion relations in the medium for the propagating states, we need to find the complex zeroes of $\mathcal{S}^{-1}$. These are given, to leading order in the mixing angle $\theta$ and neglecting light neutrino masses, by:
\begin{equation}
    \begin{gathered}
        p_0^2-p^2 + \Omega^h - \theta^2 m_{DM}^2 \frac{1}{1+m_{DM}^2/\Omega^h}=0,\quad \text{for light neutrinos},\\
        p_0^2-p^2-m_{DM}^2-\theta^2 m_{DM}^2\left(\frac{1}{\alpha^h}+\frac{1}{1+\Omega^h/m_{DM}^2}\right)=0,\quad \text{for keV DM}.
    \end{gathered}
    \label{eq:dispersion_toy}
\end{equation}
From the imaginary part of $p_0$ we can directly obtain the relaxation rate of the DM, which is given by
\begin{equation}
    \Gamma^h_s=2\theta^2 \xi \left[\frac{\mathrm{Im}\left(\alpha^h\right)}{|\alpha^h|^2}+\frac{\gamma^h/\xi}{\left(1+\Delta^h/\xi\right)^2+\left(\gamma^h/\xi\right)^2}\right],
    \label{eq:Production_2x2}
\end{equation}
where $\xi\equiv m_{DM}^2/(2p)$ and the rate depends on the helicity of the states, $h$, through the dependence of $\alpha^h$ and $\Omega^h\equiv 2p\left(\Delta^h+i \gamma^h\right)$ on it (see Eq.~(\ref{eq:self_energy_helicity})). Although we are not interested in contributions to the DM production rate beyond those involving a SM boson, for completeness we show in Appendix~\ref{app:scalar-production} the resulting production rate within this formalism when including a direct coupling between the singlet neutrinos and a new scalar, which has been extensively studied in the literature~\cite{Merle:2013wta,Fernandez-Martinez:2021ypo,Yaguna:2011qn,Silva-Malpartida:2023yks}. The expression from Eq.~(\ref{eq:Production_2x2}) agrees with that of Ref.~\cite{Lello:2016rvl} for Dirac neutrinos, which would translate into $\alpha^h\rightarrow 1$, and in which the so-called ``effective mixing angle'', $\theta_{eff}^h$, is introduced as
\begin{equation}
    \theta_{eff}^h=\frac{\theta}{\sqrt{\left(1+\Delta^h/\xi\right)^2+\left(\gamma^h/\xi\right)^2}}.
    \label{eq:effective_mixing_2x2}
\end{equation}

\begin{figure}
    \centering
    \includegraphics[width=0.4955\textwidth]{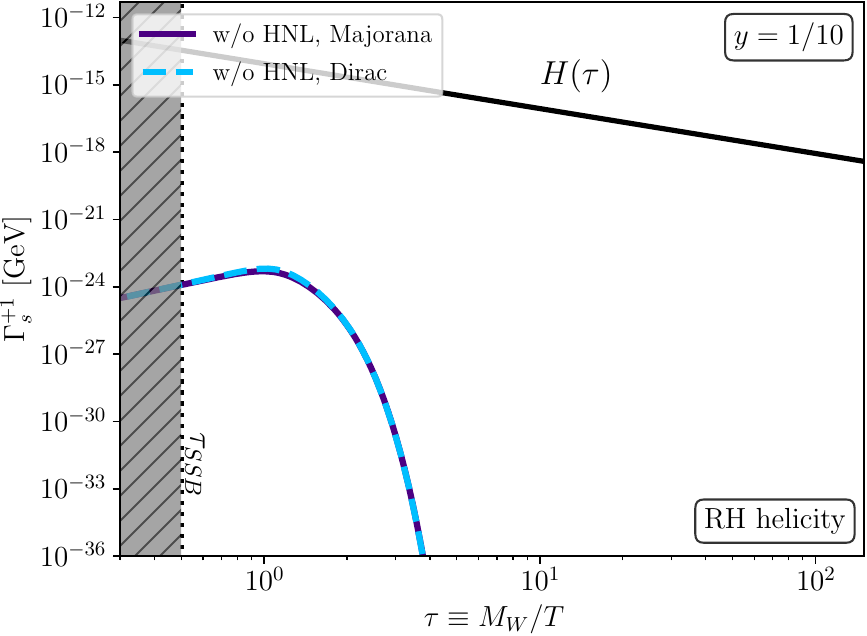}
    \includegraphics[width=0.4955\textwidth]{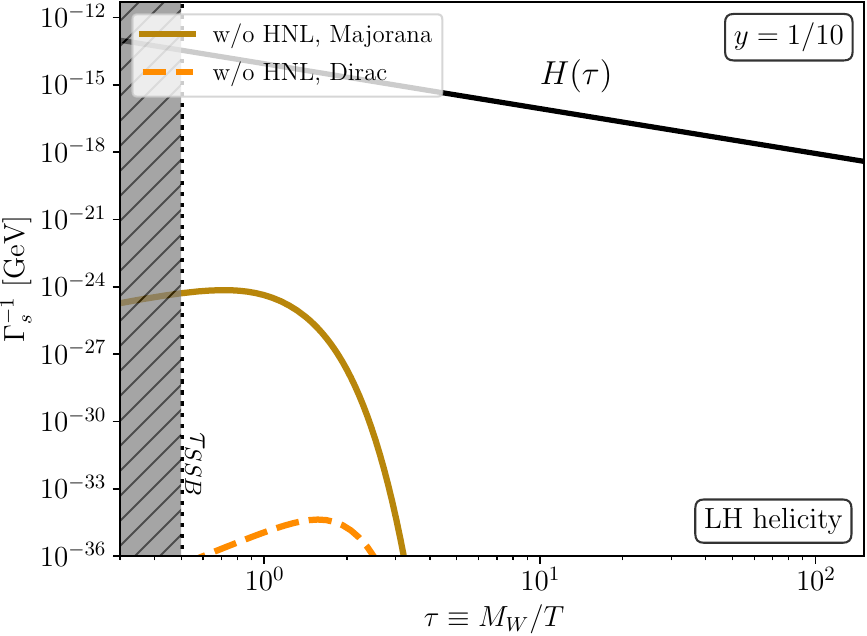}
    \caption{DM production rates for RH helicity (left panel) and LH helicity (right panel) neutrinos in the ``toy $2\times2$'' scenario. Solid lines are for Majorana neutrinos while dashed ones for Dirac neutrinos. The momenta has been fixed such that $y\equiv p/T=1/10$. The solid black line represents the Hubble expansion rate as a function of $\tau\equiv M_W/T$, while the gray shaded area represents temperatures above the EW crossover for which the results do not apply. For the neutrino parameters we fix $m_{DM}=10$~keV and $|\mathcal{U}_{\alpha 4}|\sim 10^{-6}$.}
    \label{fig:Production_rates_DiracvsMajorana}
\end{figure}
We show in Fig.~\ref{fig:Production_rates_DiracvsMajorana} the DM production rates through weak boson decays for positive helicity (left panel) and negative helicity (right panel) for Dirac neutrinos (dashed lines) and Majorana neutrinos (solid lines). The active-DM mixing is set to $|\mathcal{U}_{\alpha 4}|\sim 10^{-6}$ and $m_{DM}=10$~keV. For RH helicity neutrinos, the term proportional to $\alpha^h$ does not contribute considerably to the production rate as it is suppressed through the momenta of the neutrino, which is of the order of the temperature. Only very low momenta DM would have a non-negligible contribution, but the probability of producing such a neutrino is suppressed by the distribution function. 
Thus, most of the production of RH helicity DM goes through the second term in Eq.~(\ref{eq:Production_2x2}), which depends on $\Omega^h$. This second term is the only one present for Dirac neutrinos, and thus both rates are qualitatively similar in the left panel. On the other hand, LH helicity neutrinos present a completely different production rate depending on whether they are of Majorana or Dirac  Nature. The outstanding difference arises because, when neutrinos are Dirac particles, only the second term in Eq.~(\ref{eq:Production_2x2}) is present. In the SM, weak interactions only couple to LH neutrinos, which receive large self-energy corrections, as can also be noted from the first line of Eq.~(\ref{eq:dispersion_toy}), where $\Omega^h$ appears at leading order playing the role of a ``thermal mass''. This in turn translates into a suppression of the effective mixing angle defined in Eq.~(\ref{eq:effective_mixing_2x2}) which itself suppresses the production rate (dashed orange line in the right panel of Fig.~\ref{fig:Production_rates_DiracvsMajorana})\footnote{We have checked that this suppression can be as large as ten orders of magnitude when comparing the mixing angles in vacuum and at $T\sim M_W$.}. On the contrary, if neutrinos are Majorana particles, then the term $\alpha^h$ is present thanks to the $Z$ boson couplings to RH neutrinos, and the production rate for LH helicity DM is greatly enhanced, and qualitatively similar to the one for RH helicity DM (although not equal because of the different $W$-boson contribution). 

\section{Results and discussion}\label{Sec:analysis}

\subsection{Decay rates for different contributions}\label{Sec:rates}
\begin{figure}
    \centering
    \includegraphics[width=0.495\textwidth]{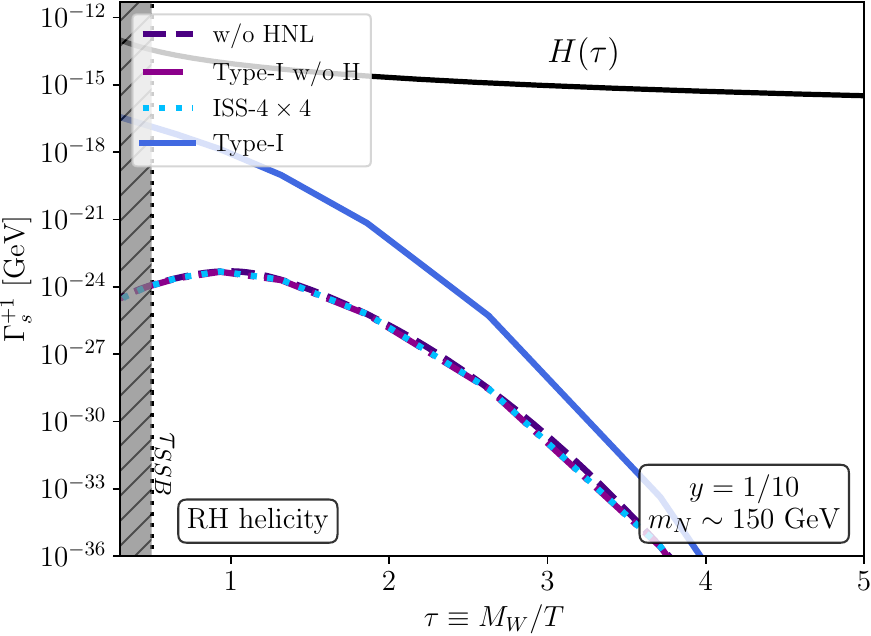}
    \includegraphics[width=0.495\textwidth]{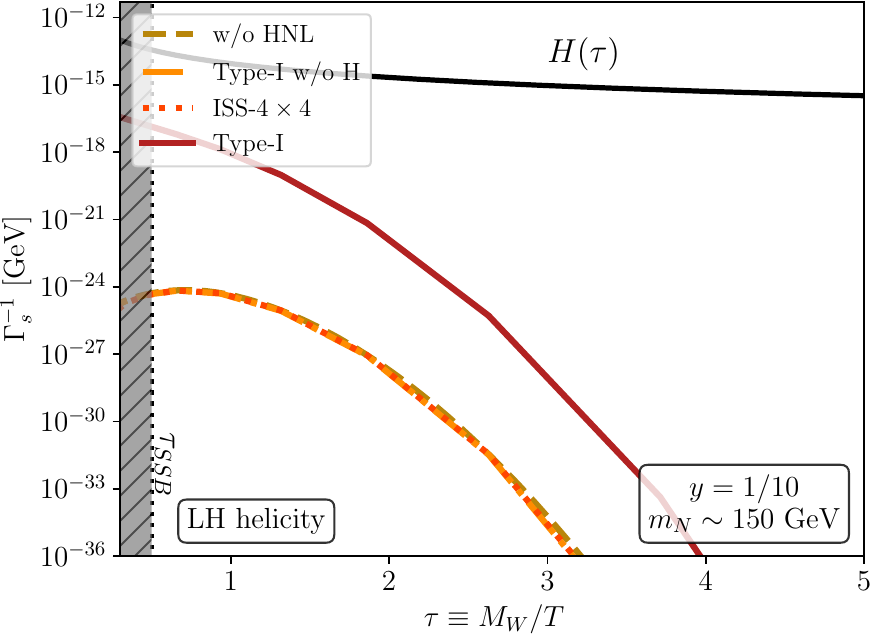}
    \caption{Production rates in the ``toy $2\times2$'' scenario without HNL (dashed lines), the ``ISS-like'' scenario (dotted lines) and the type-I seesaw (solid lines) for RH helicity neutrinos (left panel with cold hues) and LH helicity neutrinos (right panel with warm hues). Additionally, dash-dotted lines represent the results from the type-I seesaw without the contribution from the Higgs. The vertical black dotted line represents the temperature at which SSB takes place and above which the results do not apply. The solid black line represents the Hubble expansion rate. The momenta is fixed such that $y\equiv p/T=1/10$, and the mass of the HNLs, when present, is approximately given by $m_N$. The active neutrino-DM mixing is set to $|\mathcal{U}_{\alpha 4}|\sim 10^{-6}$ and the DM mass to $m_{DM}\sim 10$~keV.}
    \label{fig:Production_rates_tau}
\end{figure}

We start our discussion by studying the DM production rate, $\Gamma^h_s$, as a function of the temperature through $\tau\equiv M_W/T$, and subsequently as a function of momenta through $y\equiv p/T$, variables which will prove useful later on to estimate the total DM abundance that can be generated through this mechanism. Throughout all our discussion, we set the mass of the DM candidate to $m_{DM}=10$~keV, its mixing with active neutrinos to $\left|\mathcal{U}_{\alpha 4}\right|\sim 10^{-6}$, and the mass of the heavy neutrinos, when present, to $m_N\sim 150$~GeV, unless otherwise specified. Additionally, whenever heavy neutrinos are present, we set their mixings with active neutrinos such as, at most, they saturate current bounds on them~\cite{Blennow:2023mqx}.

In Fig.~\ref{fig:Production_rates_tau} we show the production rates for both RH helicity ($h=+1$, left panel using cold hues) and LH helicity ($h=-1$, right panel on warm hues) DM as a function of the temperature, for the different cases we have motivated in Section~\ref{sec:cases} and summarized in Table~\ref{tab:summary_scenarios}. These were:
\begin{itemize}
    \item ``Toy $2\times2$'' scenario where only weak gauge boson decays to DM are relevant (dashed lines labelled ``w/o HNL''), given that the Higgs boson couples through neutrino masses, which are negligible in comparison with the other scales of the problem lying around the EW scale.
    \item Type-I seesaw, but only taking into account weak gauge boson contributions (dash-dotted lines identified as ``Type-I w/o H''), to assess the impact of the possible heavy neutrino decays into a gauge boson and DM.
    \item ``ISS-like'' scenario where heavy neutrinos are present (dotted lines called ``ISS-$4\times4$'') and thus the Higgs contribution needs to be taken into account for consistency. We work in the single family approximation, which as discussed in Section~\ref{sec:minimal_frame},  translates into a suppression of the couplings.
    \item Type-I including all contributions, notably the Higgs decay channel (solid lines labelled ``Type-I'').
\end{itemize}
Whenever we compare the ``ISS-like'' construction with the type-I, we set the heavy neutrino scale and mixings to similar values in order to be able to extract meaningful conclusions.

The black solid line represents the Hubble expansion rate as a function of $\tau$, which serves to compare and verify that, indeed, $\Gamma_s^h(\tau)\ll H(\tau)$ and thus considering freeze-in production of DM through decays is well justified. Finally, the gray shaded area for small values of $\tau$ (larger temperatures) represents the temperature above the EW crossover at which the Higgs vev becomes zero. For temperatures above $T_{SSB}\sim 160$~GeV, we are in the symmetric phase where the results of the present study cannot be applied. In this regime production would rather go through $2\rightarrow2$ scatterings, but its contribution is beyond the scope of this work. 
Moreover, we assume the transition from the symmetric to the broken phase to be instantaneous. Thus, we consider for the Higgs and gauge bosons their $T=0$ masses, leaving the inclusion of the full temperature dependence of the Higgs vev and thermal masses in the propagators for future work. Indeed, the temperature dependence of the Higgs vev would translate into a suppression of  the gauge bosons and  Higgs masses  for temperature regimes close to $T_{SSB}$. In this case, the inclusion of the Higgs and gauge boson thermal masses in the high temperature limit (HTL) would be necessary in order to have a consistent description of the production rate. The thermal mass of the Higgs, given by
\begin{equation}
    \Pi_{\phi}=\frac{T^2}{48}\left[9g^2+3g^{\prime2}+2\left(6 Y_t^2+12\lambda+2\mathrm{tr}(Y_{\nu}^{\dagger}Y_{\nu})\right)\right],
\end{equation}
becomes particularly relevant in this regime, being the dominant contribution to the scalar mass and regularizing IR divergences that would appear otherwise~\cite{Laine:2013lka,Ghiglieri:2016xye,Ghiglieri:2021vcq}. We have nonetheless checked that, under the assumption of the instantaneous phase transition, the inclusion of the Higgs thermal mass does not qualitatively modify our results. Moreover, we could consider as well the inclusion of thermal masses for the HNL, which can be estimated as
\begin{equation}
    \Pi_{N_i}\sim \left(Y^{\dagger}_{\nu}Y_{\nu}\right)_{ii}T^2.
\end{equation}
On the other hand, there are numerous experimental constraints on the size of the mixing between active and heavy neutrinos, $\Theta$, which can be translated into bounds on the size of the Yukawa couplings as
\begin{equation}
    \left(Y^{\dagger}_{\nu}Y_{\nu}\right)_{ii}\lesssim \frac{2M_i^2}{v_H^2}\left(\Theta^{\dagger}\Theta\right)_{ii}^{exp}.
    \label{eq:bound_Yukawas}
\end{equation}
Using the result from Eq.~(\ref{eq:bound_Yukawas}) in the expression of the thermal mass, we find an upper bound on its size
\begin{equation}
    \Pi_N\lesssim \frac{2M_i^2}{v_H^2}\left(\Theta^{\dagger}\Theta\right)_{ii}^{exp} T^2.
\end{equation}
When compared with $M_i^2$, for the temperatures and masses we have investigated, and given the bounds $\left(\Theta^{\dagger}\Theta\right)_{ii}^{exp}\lesssim 10^{-1}-10^{-2}$, we find that thermal masses for the HNL are usually smaller than the actual vacuum mass.

The first feature we can observe by comparing both panels is that, in all instances, the production rates for both helicities are qualitatively similar (although not quantitatively). This stems from the fact that we have Majorana neutrinos, and thus, both $Z$ and Higgs contribute to both helicities, while interactions with the $W$ boson only contribute to the LH helicity neutrinos.

Second, we note that in all cases, except when including the Higgs contribution in the type-I seesaw like construction, the rates are qualitatively very similar, peaking around $\tau=M_W/T=1$ with values $\mathcal{O}\left(10^{-24}\right)$~GeV. It is thus clear that, without the Higgs contribution, the presence of the heavy neutral leptons does not alter significantly the dynamics in the early Universe, and thus the amount of DM generated through decays involving gauge bosons, to be quantified in the next section, is expected to be negligible regardless of the neutrino mass model details and spectrum beyond the presence of DM and its mixing with active neutrinos. 

On the other hand, focusing now on the solid lines in which we consider a type-I scenario and include all relevant contributions, the production rate gets considerably enhanced compared to the others, while still satisfying $\Gamma_{s}^h(\tau)\ll H(\tau)$. The most remarkable difference here is with the ``ISS-like'' construction, which has a very similar spectrum, but differs in the structure of the Yukawa couplings between singlet neutrinos and the Higgs, thus producing a negligible contribution.
The reason for this suppression is that, in order to satisfy the bounds on the DM-active neutrinos mixing angle, the full ISS(2,3) model requires a hierarchy of values in the mass terms coupling right-handed neutrinos and sterile singlets~\cite{Abada:2017ieq}; it is clear that such a structure cannot be achieved in a simplified model, where the only way to comply with experimental constraints is to suppress the overall Yukawa scale, thus also suppressing the HNL-Higgs couplings. We stress that the ``ISS-like'' construction analysed here is a simplified model, and we leave the study of the full (more complex) (2,3) ISS for a future work.
In particular, we find that the only way to reproduce a similar spectrum as in the full type-I with our single family approximation for the ISS is in general through a suppression of either the active-DM coupling or the active-heavy mixing, if not both, and thus the rates do not differ notably from the case without HNL. On the contrary, with the full type-I structure, cancellations can lead to spectra satisfying all constraints while featuring large couplings. We conclude then that the single family approximation, applied in this case for the ISS, cannot completely describe all the possibilities full models describing oscillations offer, as argued in previous sections, and it is therefore necessary to work with full neutrino mass models. Note that for the full type-I seesaw scenario the rate tends to increase with temperature, to be expected from the possible Bose enhancement factor in the final state of the heavy neutrino decay $n_{h}\rightarrow H+n_{DM}$. We emphasise however that these results are only applicable for temperatures below the EW crossover $T_{SSB}$.

\begin{figure}
    \centering
    \includegraphics[width=0.495\textwidth]{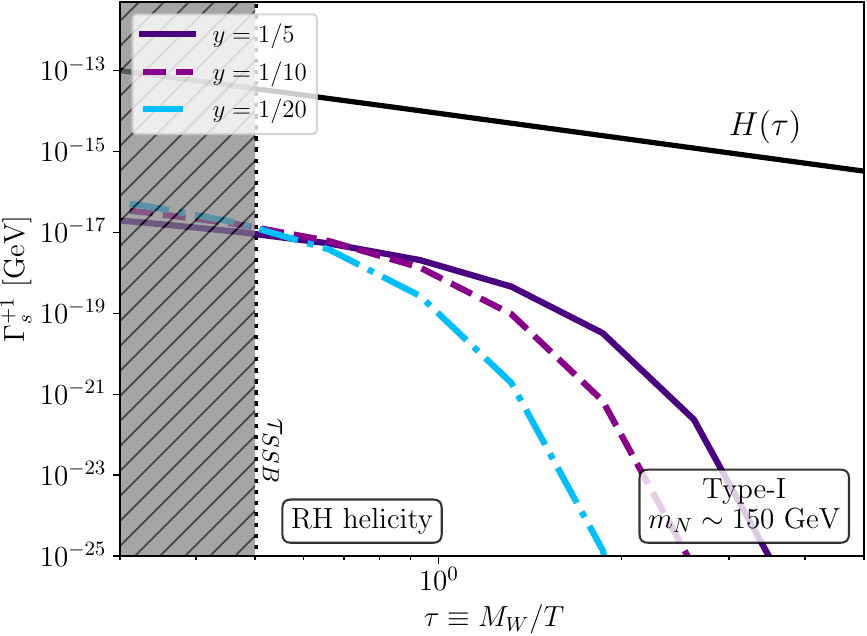}
    \includegraphics[width=0.495\textwidth]{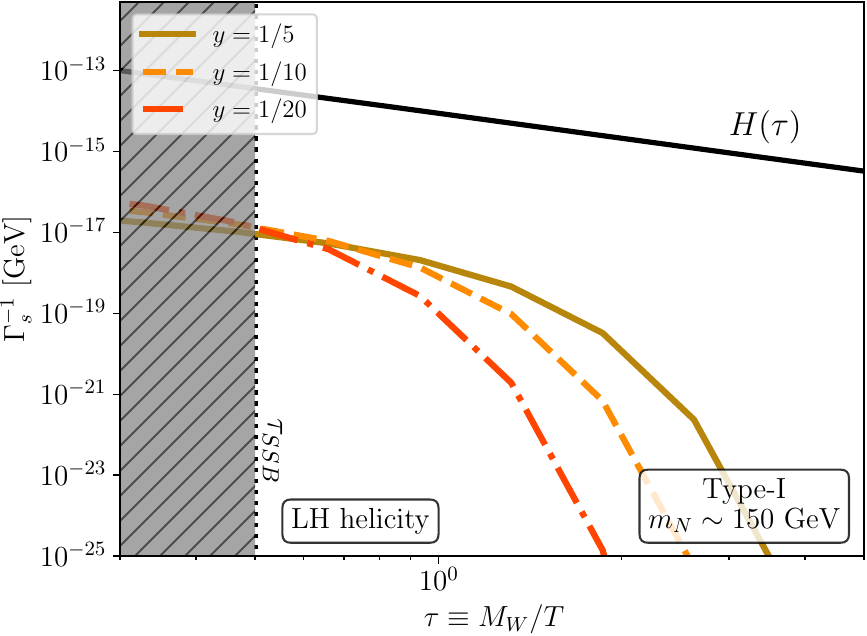}
    \caption{Production rates in the type-I seesaw for different values of $y\equiv p/T$ as a function of the temperature. The active neutrino-DM mixing is set to $|\mathcal{U}_{\alpha 4}|\sim 10^{-6}$ and the DM mass to $m_{DM}\sim 10$~keV, while the HNL mass is given by $m_N$. Left panel with cold hues shows the production for positive helicity DM while the right panel with warm hues for negative helicity. In both, the solid black line represents the Hubble expansion rate.}
    \label{fig:Production_rates_tau_typeI}
\end{figure}
We show in Fig.~\ref{fig:Production_rates_tau_typeI} the production rates, using the same color code for the different helicities as before, for several values of the ratio of momenta over temperature, $y$, for the type-I construction. For large temperatures all the rates saturate to similar values, while larger values of $y$ translate into a larger production rate at smaller temperatures. 

\begin{figure}
    \centering
    \includegraphics[width=0.495\textwidth]{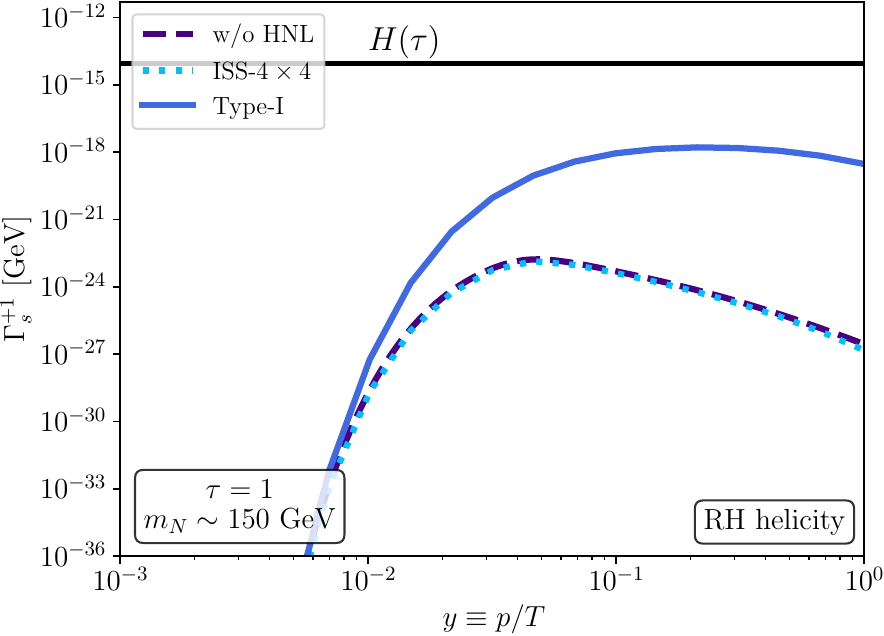}
    \includegraphics[width=0.495\textwidth]{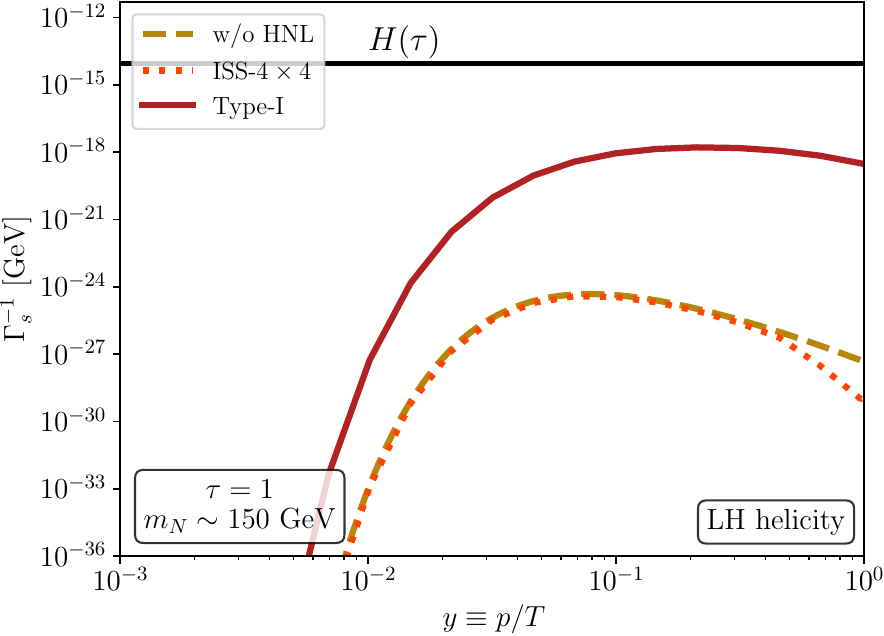}
    \caption{Production rates for positive (left panel in cold hues) and negative helicity (right panel with warm hues) neutrinos, for the case without HNL (dashed lines), the ``ISS-like'' scenario (dotted lines) and the type-I seesaw (solid lines) as a function of $y$ for fixed $\tau\equiv M_W/T=1$. The active neutrino-DM mixing is set to $|\mathcal{U}_{\alpha 4}|\sim 10^{-6}$ and the DM mass to $m_{DM}\sim 10$~keV, while the HNL mass is given by $m_N$. The black solid line corresponds to the Hubble expansion rate at the given temperature.}
    \label{fig:Production_rates_momenta}
\end{figure}

When estimating the DM abundance produced through this mechanism in the next section, we will indeed need to take into account the rate for any $y$ as well as for  $\tau\geq \tau_{SSB}$. We show in Fig.~\ref{fig:Production_rates_momenta} a comparison of the production rates as a function of $y$ between the different cases, for $\tau\equiv M_W/T=1$ fixed. It is again clear that the one family approximation for the ISS does not contribute further to the production than the ``toy $2\times 2$'' case. The peak of the production appears at $p\sim T/20$, even if in the full type-I scenario, the production rate is somewhat broader given the various decays contributing.

\begin{figure}[h]
    \centering
    \includegraphics[width=0.495\textwidth]{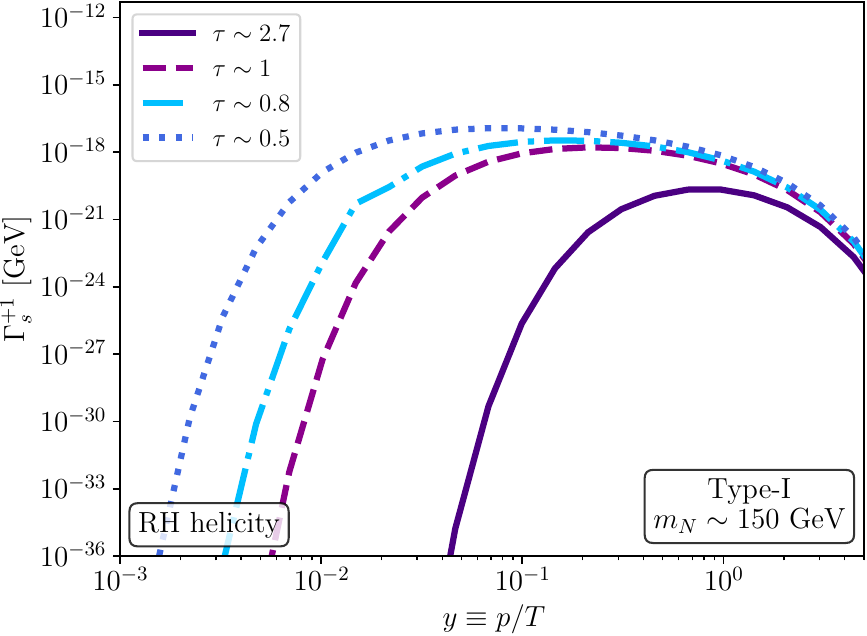}
    \includegraphics[width=0.495\textwidth]{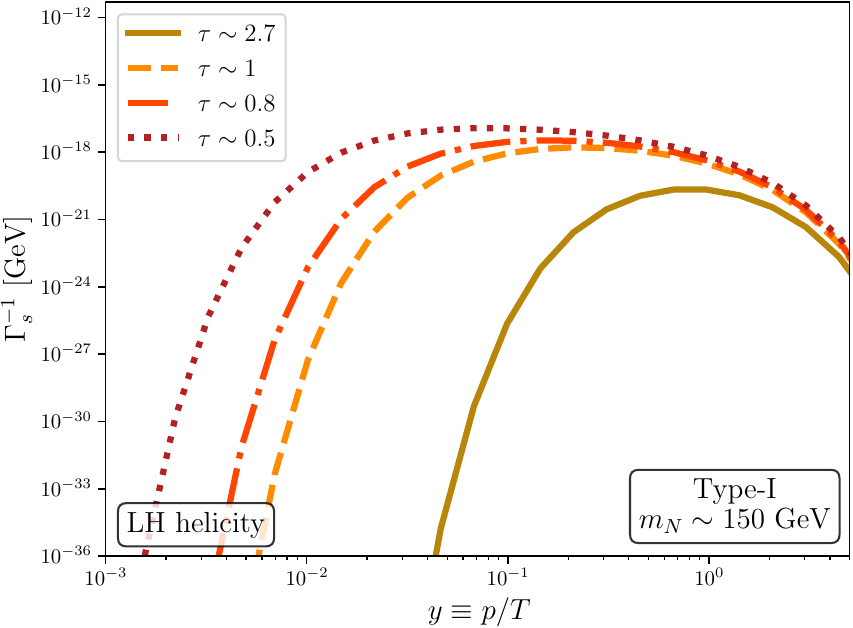}
    \caption{Production rates for positive (left panel in cold hues) and negative helicity (right panel with warm hues) DM, for the type-I seesaw, as a function of $y$ for several fixed $\tau\equiv M_W/T$. The active neutrino-DM mixing is set to $|\mathcal{U}_{\alpha 4}|\sim 10^{-6}$ and the DM mass to $m_{DM}\sim 10$~keV, while the HNL mass is given by $m_N$.}
    \label{fig:Production_rates_momenta_typeI}
\end{figure}

From the production rates in Fig.~\ref{fig:Production_rates_momenta}, and more clearly in Fig.~\ref{fig:Production_rates_momenta_typeI}, it is obvious that the probability to produce both a large or low momenta DM neutrino decreases exponentially, while around values $p\sim T/5$ the maximum of the production is achieved. Another interesting feature is the reduction of the rate as the Universe expands (larger values of $\tau$). Indeed, the decays can only happen when the heavy mother particles (mainly the heavy neutrinos and Higgs) are present in the thermal bath. At temperatures much below their masses, their abundance starts to be Boltzmann suppressed, and thus the production rate becomes negligible when the temperature of the Universe is below $T\sim m_N/15$, the usual freeze-out temperature of the mother particles.

In the next section, we will estimate the final DM abundance in the two limiting cases found here, namely the ``toy $2\times 2$'' scenario where only light neutrinos and DM are relevant, and with a benchmark point within the full type-I description which possibly renders a much larger DM abundance beyond the Dodelson-Widrow contribution through oscillations and collisions at lower temperatures~\cite{Dodelson:1993je}.

\subsection{DM production}
Following the discussion of~\cite{Lello:2016rvl}, the DM relic density can be determined by solving the following Boltzmann equation:
\begin{equation}
    \frac{d f^h_{DM}}{dt}=\Gamma^h_s (p,t) \left[f_{DM,eq}-f_{DM}^h\right],
\end{equation}
with $f_{DM}^h=f_{DM}^{\pm}$ being the distribution function of the positive and negative helicity neutrinos while $f_{DM,\rm eq}(p)={\left \{ \exp\left[p/T\right]+1 \right \} }^{-1}$ is the equilibrium Fermi-Dirac distribution for a massless particle\footnote{We are interested here in the evolution of the DM distribution function, which for the temperatures and momenta of interest can be considered to be massless for most practical purposes, and thus $E_{DM}\sim p$.}. Assuming, as customary for freeze-in scenarios, that the DM abundance is always much below the corresponding thermal equilibrium value, we can neglect the term $f_{DM}^h$ on the right-hand side of the previous equation. To account for the expansion of the Universe, one can redefine the momenta as $p\rightarrow p/a(t)$, with $a(t)$ being the cosmological scale factor. In a standard cosmological history, DM production occurs when the energy budget of the Universe is dominated by its radiation content\footnote{In some regions of the parameter space the HNL can introduce a new epoch of matter domination~\cite{Abada:2014zra}, possibly altering the standard cosmological history. We assume here this not to be the case.}, hence $a(t)=T_0/T(t)$ with $T_0$ being the present day temperature. By introducing the new variables, $y=p(t)/T(t)$ and $\tau(t)=M_W/T(t)$, the Boltzmann equation can be rewritten as:
\begin{equation}
     \frac{d f^h_{DM}}{dt}=\frac{d f^h_{DM}}{d\tau}H(\tau)\tau(t)=\Gamma^h_{\rm prod}(\tau(t),y)
     \label{eq:boltzmann}
\end{equation}
where we have conveniently defined $\Gamma^h_{\rm prod}\equiv \Gamma^h_s f_{DM,eq}$, while the Hubble parameter can be written as:
\begin{equation}
    H=1.66 g_{\rm eff}^{1/2}\frac{M_W^2}{M_{\rm pl}\tau(t)^2}
\end{equation}
in a radiation dominated Universe. The DM number density  can be computed by integrating over both $\tau$ and $y$:
\begin{align}
    & N_{\rm DM}=\sum_{h=\pm} \frac{T^3(t)}{2\pi^2}\int_0^{+\infty} n_{\rm DM}^h(y) y^2 dy,\, \textrm{with}\,
    n_{\rm DM}^h (y)=\int_{\tau_i}^{\tau_f} \frac{d f_{DM}^h}{d\tau}d\tau.
\end{align}
The fraction of DM generated through this mechanism, normalized to the observed DM relic density, can thus be written as~\cite{Lello:2015uma}:
\begin{equation}
    f_{\rm DM} = \frac{m_{\rm DM}}{7.4\,\text{eV}}{\left(\frac{\Omega_{\rm DM}^{\rm exp} h^2}{0.1199}\right)}\frac{g_{\rm DM}}{g_d}\sum_h \int_0^{+ \infty}n^h_{\rm DM}(y)y^2 dy,
\end{equation}
where $g_{DM}$ and $g_d\sim 100$ are the number of degrees of freedom for DM and the number of relativistic degrees of freedom at decoupling, respectively.

As a first estimate for the amount of DM that can be produced through freeze-in decays of heavier particles, we compute the relic density in the ``toy $2\times2$'' scenario, as it allows to efficiently integrate over both $p$ and $T$. On general grounds, even if the Majorana nature of the neutrinos translate into qualitatively different results to those found in Ref.~\cite{Lello:2016rvl}, we find good agreement with the estimates done there, finding that the produced DM abundance from weak gauge boson decays is orders of magnitude below the observed one, given the observational constraints on the neutrino mixing. Particularly, the fraction of DM that can be accounted for through freeze-in is $f_{DM}\equiv \Omega_{DM}/\Omega_{DM}^{\textrm{exp}}\sim 10^{-10}$. 

\iffalse
\begin{table}[h]
\centering
\begin{tabular}{||c |c |c |c |c ||} 
 \hline
 &  $f_{DM}^{T=0}\sim 0.14$ & $f_{DM}^{T=0}\sim 7.8$ & $f_{DM}^{T=0}\sim190$ & Thermal \\ [0.5ex] 
 \hline\hline
 $m_{DM}$~(keV) & 10 & 10 & 10 & 5 \\ 
 \hline
 $\omega_{12}$ & $10^{-7}$ & $10^{-7}$ & $10^{-7}$ & $\left(0.089+2i\right)\times 10^{-8}$ \\
 \hline
 $\omega_{13}$ & 0 & 0 & 0 & $\left(71+3.7i\right)\times 10^{-9}$ \\
 \hline
 $\omega_{23}$ & $9i$ & $10i$ & $10.8i$ & $5.5\times 10^{-10}+10.5i$ \\ [1ex] 
 \hline
\end{tabular}
\caption{DM mass and CI parameters from Eq.~(\ref{eq:complexangles}) for the four benchmark points studied in Fig.~\ref{fig:Production_rates_momenta_typeI_fraction}, identified by their DM production neglecting thermal effects. In all cases the HNL masses are $m_N\sim 150$~GeV up to small corrections, while in the last column we additionally have non-zero Majorana phases in the PMNS mixing matrix, $\alpha_1\sim 1.06\pi$ and $\alpha_2\sim 0.38 \pi$.}
\label{tab:parameters_CI}
\end{table}
\fi

\begin{figure}
    \centering
    \includegraphics[width=0.495\textwidth]{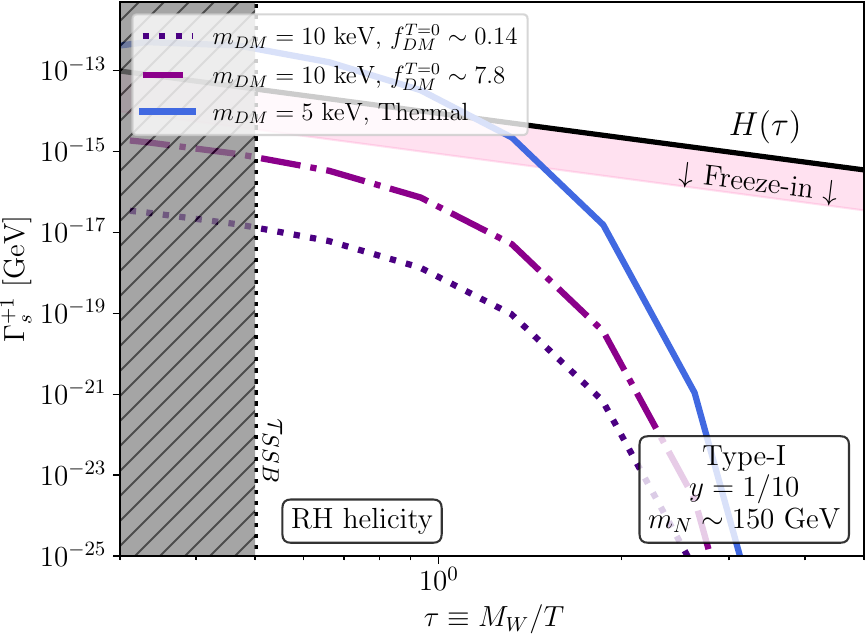}
    \includegraphics[width=0.495\textwidth]{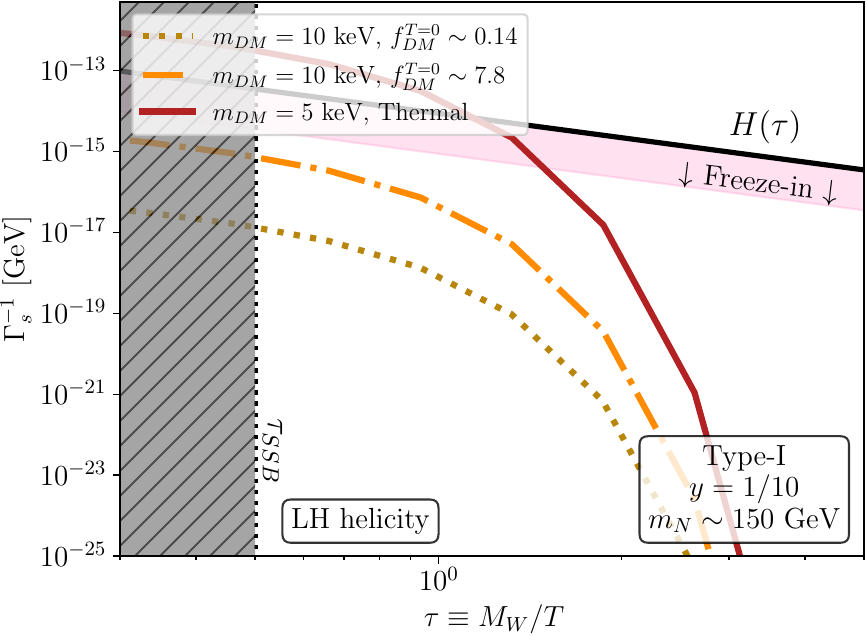}
    \caption{Production rates for positive (left panel in cold hues) and negative helicity (right panel with warm hues) DM, for the type-I seesaw, as a function of $\tau$ for different points in parameter space. The HNL mass is always set around $m_N\sim150$~GeV, while the CI parameters change as showin in Table~\ref{tab:parameters_CI}, translating into slightly different rates. The DM mass is set to $5$~keV for the solid lines while $m_{DM}=10$~keV for the rest. The solid black line represents the Hubble expansion rate.}
    \label{fig:Production_rates_momenta_typeI_fraction}
\end{figure}

\begin{table}
\centering
\begin{tabular}{||c |c |c |c ||} 
 \hline
 &  $f_{DM}^{T=0}\sim 0.14$ & $f_{DM}^{T=0}\sim 7.8$ & Thermal \\ [0.5ex] 
 \hline\hline
 $m_{DM}$~(keV) & 10 & 10 & 5 \\ 
 \hline
 $\omega_{12}$ & $10^{-7}$ & $10^{-7}$ & $\left(0.089+2i\right)\times 10^{-8}$ \\
 \hline
 $\omega_{13}$ & 0 & 0 & $\left(71+3.7i\right)\times 10^{-9}$ \\
 \hline
 $\omega_{23}$ & $9i$ & $10i$ & $5.5\times 10^{-10}+10.5i$ \\ [1ex] 
 \hline
\end{tabular}
\caption{DM mass and CI parameters from Eq.~(\ref{eq:complexangles}) for the four benchmark points studied in Fig.~\ref{fig:Production_rates_momenta_typeI_fraction}, identified by their DM production neglecting thermal effects. In all cases the HNL masses are $m_N\sim 150$~GeV up to small corrections, while in the last column we additionally have non-zero Majorana phases in the PMNS mixing matrix, $\alpha_1\sim 1.06\pi$ and $\alpha_2\sim 0.38 \pi$.}
\label{tab:parameters_CI}
\end{table}
On the other hand, as extensively discussed in Section~\ref{Sec:rates}, full neutrino mass models contribute to the production through decays of the HNL to DM and the Higgs. This can be further observed in Fig.~\ref{fig:Production_rates_momenta_typeI_fraction} where we show the rates for different points in parameter space for the type-I seesaw with a HNL scale $\mathcal{O}(150)$~GeV. The solid lines correspond to $m_{DM}=5$~keV, while the rest have masses $m_{DM}=10$~keV, with different complex angles in the CI parametrisation, specified in Table~\ref{tab:parameters_CI}. From here we note that, while complying with all bounds\footnote{The parameter points reported in Fig.~\ref{fig:Production_rates_momenta_typeI_fraction} comply with theoretical constraints and available experimental data, including: neutrino oscillation data~\cite{Esteban:2020cvm}, direct searches of HNL~\cite{L3:2001zfe,Das:2014jxa,Klinger:2014vdo,Antusch:2015mia,CMS:2018iaf}, non-unitarity bounds~\cite{Fernandez-Martinez:2016lgt,Blennow:2023mqx}, neutrinoless-double beta decay searches~\cite{Shirai:2017jyz}, X-ray bounds on decaying DM for the considered mass values~\cite{Boyarsky:2005us,Boyarsky:2006fg,Watson:2006qb,Yuksel:2007xh,Loewenstein:2008yi,Riemer-Sorensen:2009zil,Mirabal:2010an,Essig:2013goa,Horiuchi:2013noa,Riemer-Sorensen:2014yda,Tamura:2014mta,Riemer-Sorensen:2015kqa,Neronov:2016wdd,Perez:2016tcq,Ng:2019gch,Roach:2019ctw,Foster:2021ngm} and Lyman-$\alpha$ bounds for the fraction of DM produced via the DW mechanism~\cite{Baur:2017stq,Ballesteros:2020adh}.} we can find rates ranging from a fully thermalised species (solid lines) which would freeze-out ($\Gamma_s^h>H$) to production fully in the  freeze-in regime satisfying $\Gamma^h_s\ll H$. The red band roughly represents the interphase in which freeze-in stops being a complete description of the problem and neglecting the build-up of the DM abundance is no longer justified. This is given by the conservative assumption that, for rates satisfying $\Gamma^h_s\leq H/10$, freeze-in is a complete description. Note that these rates change with the momenta $p$, so that, depending on this, one might be in the freeze-in or freeze-out regime for different regions of the distribution function at different temperatures. Nonetheless taking  all this into account is beyond the scope of this work, and thus we estimate the final DM abundance within freeze-in for points of parameter spaces that satisfy $\Gamma^h_s<H$, leaving a full scan of parameter space and complete treatment of the Boltzmann equations for future work. Upon integrating Eq.~(\ref{eq:boltzmann}) we find in general at most a 2-order of magnitude reduction of the final abundance. For instance, for the CI parameters corresponding to the dotted lines in Fig.~\ref{fig:Production_rates_momenta_typeI_fraction} giving a DM fraction of around 14\% when neglecting thermal effects, we find that $f_{DM}\gtrsim 0.4$\%\footnote{The integration is numerically very expensive and thus we have performed it over a very coarse grid over $\tau$ and $y$, such that we quote a lower bound on the minimum DM fraction that can be generated when including the thermal effects.}, while the dash-dotted line produces, when including the thermal effects here studied, $f_{DM}\gtrsim 1$\%. Given that there are regions of parameter space ranging from freeze-in to  rates as large as to thermalise, we conclude that it is interesting to thoroughly study these cases, but leave the full and precise  solution of the Boltzmann equations, as well as the scan of the parameter space for future work.

\section{Conclusions}\label{sec:conclusions}
Despite all experimental and observational efforts to find and understand the nature of dark matter in the last decades, we still lack a compelling signal in any one of the current  searches other than gravitational evidence. It is therefore interesting to consider DM candidates beyond the WIMP paradigm, which might not interact with our direct detection experiments.

In this context, it proves extremely appealing to consider a joint origin for the DM as well as for light neutrino masses. Indeed, under the general assumption that neutrino masses are generated through the introduction of singlet RH neutrinos, whose Majorana mass could lie at any scale as it is not set by EW symmetry, a mostly singlet neutrino, at or around the keV-scale, would represent a good DM candidate. Given its small couplings due to mixing suppression, the production of such a DM candidate would in general be non-thermal, helping also to produce a colder spectrum to comply with structure formation bounds.

The most minimal scenario, accounting just for the DM and its mixing with active neutrinos, produces DM through neutrino oscillations and collisions which build up a large enough DM population, while keeping it always out of equilibrium in the early Universe. This is the well-known Doddelson-Widrow mechanism, which however cannot account for all the DM given the bounds from X-rays. It is therefore necessary to account for DM through another 
production mechanism, for example that of freeze-in through particle decays. 

While freeze-in production has been extensively studied before through the introduction, for example, of scalars directly coupled to the DM, here we have instead studied DM freeze-in through SM boson and HNL decays to DM. As soon as one introduces the DM candidate mixing with active neutrinos, weak gauge bosons are subject to decay to DM and one SM lepton. Moreover, the presence of heavier singlet neutrinos is necessary once neutrino oscillations are to be explained, automatically making Higgs and the HNL decays to DM themselves relevant for the DM production. Compared to the introduction of new scalars coupled in the dark sector, these channels represent an irreducible contribution to the final DM abundance on top of the DW mechanism, which becomes active at much colder temperatures, and thus 
it is necessary to understand whether enough DM can be produced or not through decays to DM involving a SM boson. For this purpose, there has been  a number of works studying some of these channels separately.  

On the one hand, Refs.~\cite{Abada:2014zra,Lucente:2021har} focused on DM production through HNL decays to the Higgs boson and the DM. The advantage of this channel is that it is not directly related to the heavily constrained active-DM mixing, thus potentially decoupling the DM production from the DM decay ones allowing to search for it. Contrary, the case for weak-boson decays to DM and either a charged lepton or a mostly active neutrino was studied in Ref.~\cite{Datta:2021elq}. While in both instances the observed DM abundance could be successfully generated, important thermal corrections were neglected~\cite{Lello:2016rvl}. 

While the weak boson decays to DM could not generate a large enough DM abundance due to the large thermal corrections that SM neutrinos receive at EW temperatures, it was necessary to study how the presence of the heavy neutrinos, and consequently the introduction of the Higgs boson mediated channel, might alter this picture. Just from naive arguments, the possible decay of the HNL to a boson and the DM can potentially take advantage of Bose enhancement in the final state, in contrast to the expected Fermi blocking in the final states from scalar decays into a pair of fermions. 

In this work, we have studied, for the first time, the production of a keV-scale neutrino DM species taking into account the dominant thermal effects when including HNL as well, extending over the work of Ref.~\cite{Lello:2016rvl}. While we find indeed that weak boson decays do not contribute further to the DW mechanism, given their very large thermal corrections suppressing active-DM mixing at large temperatures, the Higgs channel still represents an interesting avenue which needs further exploration. 

Among the different scenarios we consider, increasing in complexity from a scenario with only DM and an active neutrino to describing neutrino oscillations and DM within the type-I seesaw, we find that simplified scenarios working in the single family approximation,  where only one of each type of neutrino is introduced (namely an active neutrino, the DM one, and a HNL) do not produce a large DM abundance beyond weak boson decays, even in the presence of HNL which could decay to the Higgs boson. This is because in these cases, given the bounds on active-heavy neutrino mixing and on active-DM mixing from X-rays searches, couplings always need to be suppressed and thus production rates are negligible. On the contrary, working in full models completely accounting for oscillation data opens the door to a non-negligible production of DM through freeze-in. In these cases, one can still find solutions complying with all bounds while having  not necessarily  suppressed couplings. This is naturally realised if there is an underlying approximate lepton number symmetry, in which case light neutrino masses are also stable under radiative corrections. We explicitly verified this possibility in the type-I seesaw, finding solutions where the DM candidate even thermalises, although our focus was on regions of parameter space within the freeze-in regime. We studied several benchmark points, and found that the final DM abundance is, at most, suppressed by about three orders of magnitude with respect to the expectation when neglecting thermal effects. Given that there are regions of parameter space where an overabundance of DM can be found without thermal effects, we deem it promising to make a detailed study of this production mechanism with the thermal effects.

We leave a full scan of the parameter space for future work, given the expensive computational cost to estimate the final abundance for a single point, but motivated by the results we found here, where non-negligible rates for DM production are possible thanks to the presence of HNL and the Higgs boson in the plasma at EW temperatures. Among possible improvements that can also be taken into account, we highlight the introduction of thermal masses for the SM bosons (although we do not expect a qualitative change of the conclusions here drawn), as well as the Higgs vacuum expectation value evolution which would open the window to considering DM production beyond the SSB temperature. In this context, it is also necessary to include other production channels, namely $2\rightarrow2$ scatterings, as well as to consistently describe the LPM effect for temperatures above the EW crossover, in order to have a complete description of the production of the DM at all temperatures.

As already stressed, figuring out the contribution to DM from these channels, taking into account the dominant thermal effects below the EW crossover, is of relevance given that whenever considering keV neutrino DM and the origin of neutrino masses in realistic models, both the DW contribution as well as this contribution from HNL decays to DM and the Higgs would contribute and generate the minimal DM within the model. Whether these two mechanisms working at different temperatures can explain the observed DM abundance or not would thus translate into the necessity or not of new physics in the dark sector on top of the singlet neutrinos when considering this species of DM.

\section{Acknowledgement}
We are extremely  grateful to E. Fernandez-Martinez for his valuable comments and discussions which were at the origin of this work. This project has received funding /support from the European Union’s Horizon 2020 research and innovation programme under the Marie Skłodowska -Curie grant agreement No 860881-HIDDeN and under the Marie Skłodowska-Curie Staff Exchange  grant agreement No 101086085 – ASYMMETRY. ML is funded by the European Union under the Horizon Europe's Marie Sklodowska-Curie project 101068791 — NuBridge.

\appendix
\section{Regions of support for massive particles running in the loop}\label{app:regions_support}
In general, upong integration over the loop-momentum, $q$, we find four different combinations of Dirac delta functions, which define the regions of support of the integrals, over which their contribution is non-vanishing. For a boson of mass $M_B$\footnote{For our purposes this will be either the Higgs or the Z(W)-boson mass.} running in the loop we have in general
\begin{equation}
    \left[1+f_B(r_0)-f_F(q_0)\right]
    \left[\delta(q_0-\omega_k)-\delta(q_0+\omega_k)\right]\left[\delta(r_0-\omega_B)-\delta(r_0+\omega_B)\right],
\end{equation}
which can be divided in the following different cases (taking only the $T$-dependent part):
\begin{enumerate}
    \item $\left[f_B(p_0-\omega_k)-f_F(\omega_k)\right]\delta(q_0-\omega_k)\delta(p_0-\omega_k-\omega_B)$:
    
        The second Dirac delta forces $p_0>0$ thus describing the decay $n_i\leftrightarrows n_k B$ and its inverse process. Let us obtain the regions of support for $q$ in detail in this case. We know that $\omega_B\in \left[\omega_B^-,\;\omega_B^+\right]$, which translates into
        \begin{equation}
            \sqrt{(p- q)^2+M_B^2}\leq p_0-\sqrt{q^2+m_k^2}\leq \sqrt{(p+ q)^2+M_B^2}.
            \label{eq:Regions_delta1}
        \end{equation}
        By squaring on both sides we arrive at the following equality
        \begin{equation}
            2p_0\sqrt{\tilde{q}_{\pm}^2+m_k^2}=p_0^2+\tilde{q}_{\pm}^2+m_k^2-(p\pm \tilde{q}_{\pm})^2-M_B^2,
        \end{equation}
        which can only be satisfied if
        \begin{equation}
            \pm \tilde{q}_{\pm}<\tilde{\mu}\equiv \frac{p_0^2-p^2+m_k^2-M_B^2}{2p}.
            \label{eq:Cond_delta1}
        \end{equation}
        The solution to the inequalities from Eq.~(\ref{eq:Regions_delta1}) is then found to be
        \begin{equation}
            \tilde{q}_{\pm}=\frac{\mp \tilde{\mu} \sigma\pm\sqrt{\tilde{\mu}^2\sigma -m_k^2(1-\sigma)}}{1-\sigma},\quad \sigma\equiv \frac{p^2}{p_0^2}.
            \label{eq:solution_regions}
        \end{equation}
        This is only valid if $p_0>\sqrt{\tilde{q}_{\pm}^2+m_k^2}$ and Eq.~(\ref{eq:Cond_delta1}) are satisfied, and defines the region of support for this case. One can check that in the limit of massless fermions running in the loop the result from Ref.~\cite{Lello:2016rvl} is recovered.
    \item $-\left[f_B(|p_0|-\omega_k)-f_F(\omega_k)\right]\delta(q_0+\omega_k)\delta(p_0+\omega_k+\omega_B)$:

        We need $p_0<0$ in order for the second Dirac delta to contribute, given that $\omega_k$ ($\omega_B$) are positive definite, and describe a similar process as before changing neutrinos for antineutrinos. The regions of support in this case are
         \begin{equation}
            \tilde{q}_{\pm}=\frac{\mp \tilde{\mu} \sigma\pm\sqrt{\tilde{\mu}^2\sigma -m_k^2(1-\sigma)}}{1-\sigma},\;\text{if}\; -p_0>\sqrt{\tilde{q}_{\pm}^2+m_k^2} \;\text{and}\;\pm\tilde{q}_{\pm}<\tilde{\mu}.
        \end{equation}
    \item $\left[f_B(-p_0+\omega_k)+f_F(\omega_k)\right]\delta(q_0-\omega_k)\delta(p_0-\omega_k+\omega_B)$:

        For $p_0>0$ this term would describe the process $n_k\leftrightarrows n_i B$, while for $p_0<0$ one would have $B\leftrightarrows n_k \overline{n}_i$. We have the following for the integration limits over $q$
         \begin{equation}
            \tilde{q}_{\pm}=\frac{\mp \tilde{\mu} \sigma\pm\sqrt{\tilde{\mu}^2\sigma -m_k^2(1-\sigma)}}{1-\sigma},\;\text{if}\; p_0<\sqrt{\tilde{q}_{\pm}^2+m_k^2} \;\text{and}\;
            \begin{cases}
            \pm\tilde{q}_{\pm}<\tilde{\mu} & \text{for}\; p_0>0\\
            \pm\tilde{q}_{\pm}>\tilde{\mu} & \text{for}\; p_0<0
            \end{cases}.
        \end{equation}
    \item $-\left[f_B(p_0+\omega_k)+f_F(\omega_k)\right]\delta(q_0+\omega_k)\delta(p_0+\omega_k-\omega_B)$:

        For $p_0>0$ we describe the boson decay into a pair of neutrinos $B\leftrightarrows \overline{n}_k n_i$ while for $p_0<0$ one would have $\overline{n}_k\leftrightarrows B \overline{n}_i$. Similarly to the previous cases, we find
        \begin{equation}
            \tilde{q}_{\pm}=\frac{\mp \tilde{\mu} \sigma\pm\sqrt{\tilde{\mu}^2\sigma -m_k^2(1-\sigma)}}{1-\sigma},\;\text{if}\; -p_0<\sqrt{\tilde{q}_{\pm}^2+m_k^2} \;\text{and}\;
            \begin{cases}
            \pm\tilde{q}_{\pm}>\tilde{\mu} & \text{for}\; p_0>0\\
            \pm\tilde{q}_{\pm}<\tilde{\mu} & \text{for}\; p_0<0
            \end{cases}.
        \end{equation}

\end{enumerate}

\section{Useful integrals with distribution functions}\label{app:integrals}
Throughout the computations we need to perform some integrals over the distribution functions of bosons and fermions. In particular, we make extensive use of the following ones\footnote{In the following, we introduce $\sigma$ to take into account the possibility that the fermion is not present in the plasma and thus its density distribution should not be taken into account by making $\sigma=0$. Otherwise, for particles in thermal equilibrium we have $\sigma=1$.}:
\begin{equation}
\begin{split}
    \int dq \frac{q}{\omega_k}\left[f_B(\pm p_0-\omega_k)-\sigma f_F(\omega_k)\right]=&-\frac{1}{\beta}\log{\frac{1-e^{-\beta(\pm p_0-\omega_k)}}{1+\sigma e^{-\beta \omega_k}}},\\
    \int dq \frac{q}{\omega_k}\left[f_B(\pm p_0+\omega_k)+\sigma f_F(\omega_k)\right]=&\frac{1}{\beta}\log{\frac{1-e^{-\beta(\pm p_0+\omega_k)}}{1+\sigma e^{-\beta \omega_k}}},\\
    \int dq q\left[f_B(\pm p_0-\omega_k)-\sigma f_F(\omega_k)\right]=&-\frac{\omega_k}{\beta}\log{\frac{1-e^{-\beta(\pm p_0-\omega_k)}}{1+\sigma e^{-\beta \omega_k}}}\\
    &-\beta^{-2}\mathrm{Li}_2\left(e^{-\beta(\pm p_0-\omega_k)}\right)-\sigma \beta^{-2}\mathrm{Li}_2\left(-e^{-\beta\omega_k}\right),\\
    \int dq q\left[f_B(\pm p_0+\omega_k)+\sigma f_F(\omega_k)\right]=&\frac{\omega_k}{\beta}\log{\frac{1-e^{-\beta(\pm p_0+\omega_k)}}{1+\sigma e^{-\beta \omega_k}}}\\
    &-\beta^{-2}\mathrm{Li}_2\left(e^{-\beta(\pm p_0+\omega_k)}\right)+\sigma \beta^{-2}\mathrm{Li}_2\left(-e^{-\beta\omega_k}\right).
\end{split}
\end{equation}

\section{Example of production rate with new scalars}\label{app:scalar-production}
While we have extensively discussed SM and HNL contributions to the DM production, for completeness, we consider here the case in which the dark sector is comprised of other species, for example a singlet scalar, which could be at the source of the DM mass. This could be motivated by the assumption that lepton number is not explicitly broken in a seesaw scenario but rather dynamically when a singlet scalar develops a vev. On general grounds, we can have the following Lagrangian\footnote{Here we just focus on the active and DM sectors, which allows to have an analytical understanding of the results, but this can be trivially embedded into a more general setup with the heavy neutrinos as well.}
\begin{equation}
    \mathcal{L}\supset -\bar{L}_L Y_{\nu}\tilde{\Phi}\nu_S^c-\frac{1}{2}\bar{\nu}_S^cY_N \phi \nu_S+h.c.+V\left(\Phi^{\dagger}\Phi,|\phi|^2\right),
    \label{eq:lag_Scalar}
\end{equation}
where $\phi$ is a singlet scalar with lepton number $L_{\phi}=-2$. The scalar potential, $V$, is such that, upon spontaneous symmetry breaking, both the SM Higgs boson $\Phi$ and  the scalar singlet $\phi$ develop a vev, $v_H$ and $v_{\phi}$ respectively. In the mass basis, we obtain a new interaction term between the mostly singlet scalar, $\varphi$, and the neutrino mass eigenstates
\begin{equation}
    \mathcal{L}_{\varphi\nu}=-\frac{\varphi}{v_{\phi}}\sum_{i,j}\bar{n}_i\left[m_i\delta_{ij}-C_{ij}\left(m_i P_L+m_jP_R\right)-C_{ij}^*\left(m_i P_R+m_j P_L\right)\right]n_j.
    \label{eq:lag_Scalar_MassBasis}
\end{equation}
One can directly compute the contribution of such an interaction to the dispersion relations for the neutrinos, just by exchanging $v_H\rightarrow v_{\phi}$ and $\mathcal{A}_{ij}^L\rightarrow m_i-C_{ij}m_i-C_{ij}^*m_j$ in the Higgs contribution computed in Section~\ref{Sec:self-energy}. To leading order in the mixings, and neglecting the contribution from light neutrino masses, the self-energy correction arising from the mostly-singlet scalar\footnote{Note that this term gives a direct contribution to the second diagonal element in Eq.~(\ref{eq:Matrix_Inverse_Propagator}), directly related to the sterile sector.}, $\Omega^h_{\phi}\equiv 2p(\Delta^h_{\phi}+i\gamma^h_{\phi})$, contributes to the DM production rate as
\begin{equation}
    \Gamma_s^h=2\xi \left\lbrace\frac{\gamma^h_{\phi}}{\xi}+\theta^2\left(\frac{\textrm{Im}{\alpha^h}}{|\alpha^h|^2}+\frac{(\gamma^h-\gamma^h_{\phi})/\xi}{\left(1+\frac{\Delta^h-\Delta^h_{\phi}}{\xi}\right)^2+\left(\frac{\gamma^h-\gamma^h_{\phi}}{\xi}\right)^2}\right)\right\rbrace.
    \label{eq:production_rate_Scalar}
\end{equation}
It is easy to notice that, in the absence of this extra scalar interaction, we recover the original result which depends on the neutrino mixing, see Eq.~(\ref{eq:Production_2x2}). More importantly, there is now a contribution to the DM production which is not suppressed by the neutrino mixing $\theta$, and proportional to the Yukawa coupling in the dark sector, $Y_N$, thus allowing for an efficient DM production as expected~\cite{Yaguna:2011qn,Merle:2013wta,Silva-Malpartida:2023yks,Konar:2021oye}.

\bibliographystyle{JHEP}
\bibliography{Biblio}

\end{document}